\PassOptionsToPackage{unicode}{hyperref}
\PassOptionsToPackage{hyphens}{url}
\PassOptionsToPackage{dvipsnames,svgnames,x11names}{xcolor}
\documentclass[
  11pt,
]{article}
\usepackage{amsmath,amssymb}
\usepackage{iftex}
\ifPDFTeX
  \usepackage[T1]{fontenc}
  \usepackage[utf8]{inputenc}
  \usepackage{textcomp} % provide euro and other symbols
\else % if luatex or xetex
  \usepackage{unicode-math} % this also loads fontspec
  \defaultfontfeatures{Scale=MatchLowercase}
  \defaultfontfeatures[\rmfamily]{Ligatures=TeX,Scale=1}
\fi
\usepackage{lmodern}
\ifPDFTeX\else
\fi
\IfFileExists{upquote.sty}{\usepackage{upquote}}{}
\IfFileExists{microtype.sty}{% use microtype if available
  \usepackage[]{microtype}
  \UseMicrotypeSet[protrusion]{basicmath} % disable protrusion for tt fonts
}{}
\makeatletter
\@ifundefined{KOMAClassName}{% if non-KOMA class
  \IfFileExists{parskip.sty}{%
    \usepackage{parskip}
  }{% else
    \setlength{\parindent}{0pt}
    \setlength{\parskip}{6pt plus 2pt minus 1pt}}
}{% if KOMA class
  \KOMAoptions{parskip=half}}
\makeatother
\usepackage{xcolor}
\usepackage[margin=1in]{geometry}
\usepackage{longtable,booktabs,array}
\usepackage{calc} % for calculating minipage widths
\usepackage{etoolbox}
\makeatletter
\patchcmd\longtable{\par}{\if@noskipsec\mbox{}\fi\par}{}{}
\makeatother
\IfFileExists{footnotehyper.sty}{\usepackage{footnotehyper}}{\usepackage{footnote}}
\makesavenoteenv{longtable}
\usepackage{graphicx}
\makeatletter
\def\maxwidth{\ifdim\Gin@nat@width>\linewidth\linewidth\else\Gin@nat@width\fi}
\def\maxheight{\ifdim\Gin@nat@height>\textheight\textheight\else\Gin@nat@height\fi}
\makeatother
\setkeys{Gin}{width=\maxwidth,height=\maxheight,keepaspectratio}
\makeatletter
\def\fps@figure{htbp}
\makeatother
\ifLuaTeX
  \usepackage{luacolor}
  \usepackage[soul]{lua-ul}
\else
  \usepackage{soul}
\fi
\ifLuaTeX
  \usepackage{selnolig}  % disable illegal ligatures
\fi
\IfFileExists{bookmark.sty}{\usepackage{bookmark}}{\usepackage{hyperref}}
\IfFileExists{xurl.sty}{\usepackage{xurl}}{} % add URL line breaks if available
\hypersetup{
  pdftitle={Agent Reliability Profiles in Financial Services},
  colorlinks=true,
  linkcolor={Maroon},
  filecolor={Maroon},
  citecolor={Blue},
  urlcolor={Blue},
  pdfcreator={LaTeX via pandoc}}

\title{Agent Reliability Profiles in Financial Services}
\author{
Mike Hsu\textsuperscript{1} \quad
Medha Bankhwal\textsuperscript{2} \quad
B\'{e}atrice Moissinac\textsuperscript{3} \\[2pt]
Kevin Werbach\textsuperscript{4} \quad
Lukasz Szpruch\textsuperscript{5} \quad
Bennett Hillenbrand\textsuperscript{6}
\\[10pt]
\small
\begin{tabular}{@{}c@{}}
\textsuperscript{1}~MLCommons; Cambridge Judge Business School; former Acting Comptroller of the Currency \\
\textsuperscript{2}~MLCommons; Mezuro \\
\textsuperscript{3}~Zendesk \\
\textsuperscript{4}~University of Pennsylvania Wharton School \\
\textsuperscript{5}~University of Edinburgh, The Alan Turing Institute \\
\textsuperscript{6}~MLCommons
\end{tabular}
}
\date{September 2026 \\ \emph{Working draft circulated for comment.}}

\begin{document}
\maketitle

\begin{abstract}
AI agents can take actions. At times, those actions can go beyond what is intended. Agent reliability can be defined as assurance that an agent will stay within intended bounds and operate within limits. Today, there is no shared framework or language for describing, validating, and benchmarking the reliability of agentic deployments in financial services. This makes it difficult for financial institutions, vendors, and regulators to assess and trust agents at scale, thus limiting the pace of development and adoption. A standardized, shared representation of agent reliability would fill the gap. This paper introduces the Agent Reliability Profile, a per-agent unit of assurance evidence for agent deployments in financial services. Each Profile records a bounded, falsifiable claim---\emph{this agentic system reliably functions within its operating boundary.} We define ``operating boundary'' as an agent having; (1) a defined autonomy tier, (2) a defined operational design domain, (3) defined classes of action, and (4) a defined control envelope. Production assurance progresses through three levels while the Profile schema remains constant: a Profile Builder compiles a Level~1 Asserted Profile from institutional evidence, a Profile Validator tests the deployment in its own environment to produce a Level~2 Validated Profile, and operation of the same tests by a qualified independent assessor produces a Level~3 Verified Profile. Separately a Benchmarked Profile reports results comparable across institutions under reference conditions. We describe the architecture, the artifact, the assurance ladder, the comparability flag, associated tools, an evaluation methodology, applications for financial institutions and supervisors, limitations, and a staged implementation program.

\end{abstract}

\hypertarget{introduction}{%
\section{1 Introduction}\label{introduction}}

AI agents can take actions. Unlike chatbots, agents invoke tools---API calls, file writes, code commits---that can change state in the world. At times, those actions can go beyond what is intended.

We define \textbf{agent reliability} as assurance that an agent will operate within intended bounds and limits. Today each deployer defines those bounds and limits idiosyncratically. This paper proposes a common form and standardized taxonomy in which that assurance can be expressed, tested, and compared for agent deployments in financial services.\footnote{The framework and approach here could also be applied to non-financial domains, but taxonomy details, especially pertaining to the operational design domain axis, would need to be crafted by domain experts.}

AI agents are already in use at financial institutions. For instance, BNY reports that 20,000 employees are building agents on its Eliza platform, which supports more than 125 live AI use cases \href{https://openai.com/index/bny/}{\ul{{[}36{]}}}. DBS has specialized agents that draft corporate credit assessments for approximately 1,500 employees \href{https://www.dbs.com/NewsPrinter.page?locale=en\&newsId=msd3p9eh}{\ul{{[}40{]}}}. Stripe reports more than 1,300 agent-produced pull requests merged each week after human review \href{https://www.linkedin.com/posts/stripe_minions-have-changed-software-engineering-activity-7430340204831608832-xC1H}{\ul{{[}41{]}}}. As use cases expand in scope and agent adoption accelerates across the industry, agent reliability challenges are poised to increase, as discussed in the next section.

\hypertarget{the-problem}{%
\subsection{1.1 The problem}\label{the-problem}}

Risk management in financial services is fairly mature. The concepts, language, frameworks, supervisory expectations, and regulatory requirements for most risk areas are well-understood and share common foundations. Agentic AI challenges this.

Today there is no shared framework or language describing, validating, or benchmarking the reliability of agentic deployments in financial services. No published standard specifies how to state what a deployed agent is permitted to do, how autonomously, in what context, and with what containment, in a form that can be tested or independently validated. Financial institutions, vendors, independent assessors, and regulators each work with their own vocabulary. Every relying party---a counterparty, end customer, risk committee, Board, examination team---depends on the judgment, competence, and self-description of the deploying institution. Even deploying teams do not always fully measure the reach of their own agents {[}37{]}. The agentic governance and security tools currently on the market each use their own schemas, further entrenching the variability problem.

Agentic systems are also dynamic. An agent's risk profile shifts when its foundation model is updated, when its tool catalog grows, when its credentials change, and when its environment evolves. A deployment that was bounded at one moment can become unbounded at another without any explicit institutional decision. As a result, many institutions struggle to answer, in consistent terms, which agents across business units are in production, what those agents are permitted to do, and what contains them.

Regulators lack a shared vocabulary for discussing agent risk with the institutions they oversee and have limited ability to compare deployments or identify outliers across firms. In the U.S., this has been exacerbated by the interagency model risk management guidance issued by the federal banking agencies in April 2026 (SR 26-2), which places generative and agentic AI outside the scope of model risk management {[}4{]}.

Because regulators and other relying parties cannot easily assess agents at scale, trust and adoption are proceeding more slowly and unevenly than underlying capabilities warrant. If a serious harm event involving an agent at a regulated institution forces a regulatory response before a shared framework exists, that response will likely be coarsely calibrated, at material cost to beneficial deployment.

\hypertarget{objective}{%
\subsection{1.2 Objective}\label{objective}}

\emph{To establish a standardized, per-agent unit of assurance evidence---the Agent Reliability Profile---for agentic deployments in financial services, through which institutions make a bounded, testable claim about an agent's reliability.}

The answer to the question ``Is this agent reliable?'' depends on how much the agent does on its own, where it operates, what it is permitted to do, and what contains it. The Agent Reliability Profile (``Profile'') converts that open-ended question into a scoped assertion that can be tested and falsified: \emph{This agentic system reliably functions at a stated autonomy tier, within a stated operational design domain, for stated action classes, under a stated control envelope}. The construction borrows from safety engineering in other domains, such as automated driving, where the level of automation that is safe depends on operating conditions {[}10{]}. Agents can take a far wider range of actions than conveying passengers between two points, so a single conditioning axis is insufficient. The Profile uses four.

This paper makes several contributions.

First, it defines an agentic deployment's ``operating boundary'' in terms of a four axis architecture: (1) autonomy tier, (2) operational design domain, (3) action class, and (4) control envelope. This converts an open-ended reliability question into a deployment-scoped, falsifiable claim capable of credibly providing assurance to external parties (Sections~\protect\hyperlink{the-bounded-claim-and-its-formal-schema}{3} and~\protect\hyperlink{the-four-axes}{4}).

Second, it defines the Agent Reliability Profile and associated Profile Package as evidence-bearing artifacts that state: minimum performance guarantees (floors), where the evidence came from (provenance), how long the claim holds (validity windows), and what voids it (invalidation rules) (Section~\protect\hyperlink{the-agent-reliability-profile-artifact}{5}).

Third, it defines a graduated production assurance ladder of three levels, a separate benchmark comparability flag, and the tools that produce them: a Profile Builder, Profile Validator, Profile Verifier, and Profile Benchmarker (Section~\protect\hyperlink{graduated-assurance-three-levels-a-comparability-flag-and-their-tools}{6}), with an accompanying evaluation methodology (Section~\protect\hyperlink{evaluation-methodology}{7}).

Fourth, it places the Profile within the broader agent assurance ecosystem, which is comprised of several layers---product evaluation, institutional governance, identity, and runtime control. It shows how the deployment layer connects them (Section 9), and sets out applications, limitations, and a staged implementation program with defined off-ramps (Sections~\protect\hyperlink{use-in-financial-services}{8}--\protect\hyperlink{implementation-program-and-research-agenda}{11}).

The initial application for Agent Reliability Profiles is financial services, with the range of use cases covered in the discussion of operational design domains (Section \protect\hyperlink{axis-2-operational-design-domain}{4.2}).

\hypertarget{design-objectives-and-unit-of-analysis}{%
\section{2 Design objectives and unit of analysis}\label{design-objectives-and-unit-of-analysis}}

\hypertarget{the-deployment-as-the-unit-of-analysis}{%
\subsection{2.1 The deployment as the unit of analysis}\label{the-deployment-as-the-unit-of-analysis}}

The unit of the Profile is a deployed agentic system, not a vendor, product, or model release. A deployment identity includes, at minimum, the model and version, the harness and its policy configuration, the tool catalog (including MCP server and their catalogs), the credential and role configuration, the operating environment (including hosting and inference platforms of the model), the accountable owner, and the time at which this configuration was in force. Multi-agent systems must also record how the agents are wired together, how they talk to each other, who may delegate what, and how credentials pass between them. For example, if an agent can communicate with another agent via MCP or A2A protocols, the profile also includes what the 2nd agent may do on behalf of the agent being profiled.

The Profile attaches to the deployment because that is where operational risk arises and is actionable. Vendor documentation can establish product capabilities, but institutional configuration establishes what the deployed instance can access and do. For instance, an agent running in a sandboxed test environment with read-only credentials is a different risk object from the same product holding production credentials. A vendor assessment can be useful evidence inside a Profile Package, as long as issued Profiles stay bound to the deployment identity. Reusable capability catalogues of the kind proposed by Szpruch et al. {[}15{]} can also reduce assurance cost by serving as pooled inputs to multiple deployment Profiles. But each deployed combination of capabilities still needs its own Profile, because the setting and the way those capabilities work together can change what the agent can do and how it might fail.

\hypertarget{reliability-as-a-bounded-engineering-claim}{%
\subsection{2.2 Reliability as a bounded engineering claim}\label{reliability-as-a-bounded-engineering-claim}}

Reliability in this paper means performance and containment under stated conditions over a stated interval. This has four consequences.

First, scoping an agent is integral to its measurement. This requires defining the deployment boundary, domain, actions, and controls.

Second, any critical flags to agent performance are prioritized over other aggregate assessments. A failed safety floor blocks issuance of a Profile even when average task performance is high.

Third, evidence quality is included in the Profile results. Assertions, in-environment measurements, independent measurements, and cross-institution benchmarks support different claims, which are distinguished by the Profile.

Fourth, material change leads to a change in validity. Model updates, new tools, expanded credentials, and weakened controls can void or narrow a Profile.

\hypertarget{requirements}{%
\subsection{2.3 Requirements}\label{requirements}}

A usable Profile system should satisfy nine requirements:

\begin{enumerate}
\def\labelenumi{\arabic{enumi}.}
\item
  \begin{quote}
  \emph{Specificity:} identify a deployment and the exact conditions the claim covers.
  \end{quote}
\item
  \begin{quote}
  \emph{Comparability of description:} enable two institutions to describe functionally similar deployments in the same vocabulary.
  \end{quote}
\item
  \begin{quote}
  \emph{Risk-based assurance:} support low-cost inventory coverage and higher-cost assurance for higher risk or more consequential deployments.
  \end{quote}
\item
  \begin{quote}
  \emph{Evidence traceability:} permit reconstruction of field values and evaluated runs from sources and traces.
  \end{quote}
\item
  \begin{quote}
  \emph{Machine readability:} support inventories, change detection, policy checks, and supervisory analysis.
  \end{quote}
\item
  \begin{quote}
  \emph{Cryptographic integrity:} bind the issued artifact to its issuer, subject identity, evidence hashes, method version, and validity period.
  \end{quote}
\item
  \begin{quote}
  \emph{Change sensitivity:} distinguish changes that void a Profile, changes that require partial retesting, and changes with no effect on the claim.
  \end{quote}
\item
  \begin{quote}
  \emph{Software neutrality:} admit conforming implementations from institutions, assurance firms, vendors, and open-source projects.
  \end{quote}
\item
  \begin{quote}
  \emph{Compositionality:} permit a platform (Section 2.5) to carry an enforcement-claim Profile whose guarantees are inherited by reference in the Profiles of hosted agents instantiated on it, without altering the claim structure of those Profiles.
  \end{quote}
\end{enumerate}

\hypertarget{scope-and-status-of-the-profile}{%
\subsection{2.4 Scope and status of the Profile}\label{scope-and-status-of-the-profile}}

The Profile has a specific purpose and should not be relied upon beyond its intended scope.

\begin{enumerate}
\def\labelenumi{\arabic{enumi}.}
\item
  \begin{quote}
  The Profile complements, without replacing, the institutional frameworks and other agent governance mechanisms noted in Section~\protect\hyperlink{the-role-of-the-profile-in-agent-assurance}{9}.
  \end{quote}
\item
  \begin{quote}
  A Profile is evidence about an agent's configuration at a point in time.
  \end{quote}
\item
  \begin{quote}
  The Profile is software-agnostic: any tool that meets the bar and produces a conforming Profile should be acceptable, for instance as described in the reference implementations in Section~\protect\hyperlink{graduated-assurance-three-levels-a-comparability-flag-and-their-tools}{6}.
  \end{quote}
\item
  \begin{quote}
  The Profile's scope is bounded. It records operational reliability evidence, not other dimensions of workflow and model specific fairness, explainability, data governance, compliance, and cybersecurity adequacy, though Profile evidence may inform those where relevant.
  \end{quote}
\end{enumerate}

\hypertarget{platform-instantiated-deployments-and-the-platform-baseline-profile}{%
\subsection{2.5 Platform-instantiated deployments and the Platform Baseline Profile}\label{platform-instantiated-deployments-and-the-platform-baseline-profile}}

An increasing share of agent deployments in financial services are being assembled from platforms, not as discrete agentic systems. Several major U.S. financial institutions have developed shared substrates on which their teams build and govern their own agents through a control plane. For instance, Stripe's Knowledge AI platform (KAI) connects a single agent service to more than a thousand skills and tools and lets domain teams create and modify their own agents continuously {[}35{]}. BNY reports twenty thousand employees building agents on its Eliza platform under a dedicated release-approval regime {[}36{]}. Comparable first-party platforms are in production at JPMorgan, Citi, and Wells Fargo.

Such platforms are not deployments per se, at least not as defined in Section \protect\hyperlink{the-deployment-as-the-unit-of-analysis}{2.1}. Their operational design domain is open by construction, the available tools are determined per session rather than enumerated in advance, and their population of agents turns over continuously. A single Profile for a platform as a whole would be too broad to bind anything.

Platform-derived agents warrant distinct treatment, as they differ from non-platform-derived agents in three ways: (1) platform-derived agents can be created much faster and modified more easily and frequently, (2) reachable actions are set by the control plane at the platform level, so a per-agent Profile would carry little risk information (unless it included the platform controls), and (3) per-agent Profiles would not aggregate, and correlated faults --- e.g., a compromised catalog entry or an over-broad credential template --- would appear in every agent at once and in none of their Profiles.

\emph{The Platform Baseline Profile.} To address these differences, an agent builder platform should carry its own Baseline Profile, whose bounded claim is a ceiling on what sessions and derived agents can do. Specifically, for every session the platform hosts, effective autonomy should not exceed a stated ceiling without a stated gate, action classes outside a stated set should be unavailable in software rather than prohibited by policy, and no session should be able to reach data outside its authorized context. Each session\textquotesingle s effective envelope should be immutably recorded, making the claim auditable after the fact. Because the platform claim applies to all agents built on it, a single counterexample would falsify it. Assurance therefore rests on the design and construction of the platform's control plane, not by configuration discipline, with adversarial session-level testing serving as the falsification method. A Platform Baseline Profile attests to containment only. It says nothing about whether any particular agent\textquotesingle s work is good.

\emph{Profile inheritance for hosted agents}\textbf{.} An agent instantiated on a platform (``hosted agent'') carries an ordinary Profile that also records a platform anchor: the parent Platform Baseline Profile and version, incorporated by reference, whose controls form part of the agent\textquotesingle s control envelope. The hosted agent\textquotesingle s own claim is otherwise unchanged. But an anchored Profile is void if the parent claim is void --- a breach of the platform ceiling invalidates every hosted agent Profile resting on it. Because the platform\textquotesingle s control plane already records each hosted agent\textquotesingle s configuration, Level 1 anchored Profiles can be generated largely automatically.

To carry a Baseline Profile, a platform must exhibit six properties, which double as its test targets:

\begin{itemize}
\item
  \begin{quote}
  an autonomy ceiling---the maximum tier grantable to any hosted agent---enforced in software, with defined gating for any escalation,
  \end{quote}
\item
  \begin{quote}
  a closed, versioned registry of skills and tools with admission standards,
  \end{quote}
\item
  \begin{quote}
  per-session resolution and immutable recording of the effective envelope,
  \end{quote}
\item
  \begin{quote}
  technically enforced data-context isolation,
  \end{quote}
\item
  \begin{quote}
  version control over the profile's technical components, and
  \end{quote}
\item
  \begin{quote}
  change impact mapping---precise tracking of component relationships to ensure that substrate or skill updates only invalidate directly dependent Profiles (Section \protect\hyperlink{the-agent-reliability-profile-artifact}{5}).
  \end{quote}
\end{itemize}

\hypertarget{the-bounded-claim-and-its-formal-schema}{%
\section{3 The bounded claim and its formal schema}\label{the-bounded-claim-and-its-formal-schema}}

For each agentic deployment, the institution should be able to state and support the following:

\begin{quote}
\textbf{The bounded claim.} This agentic system functions reliably within its operating boundary -- i.e., at a defined autonomy tier, within a defined operational design domain, for defined classes of action, under a defined control envelope.
\end{quote}

The claim tells a relying party the conditions under which the agent may be trusted and, with equal precision, the conditions under which the claim does not apply.

Formally, let \(D\) denote a deployment. Its Profile scope is the tuple
\[
S(D) = \left( T,\mspace{6mu} O,\mspace{6mu} A^{+},\mspace{6mu} A^{-},\mspace{6mu} C \right)
\]
where \(T\) is the autonomy tier, \(O\) is the operational design domain instance, \(A^{+}\) and \(A^{-}\) are the permitted and prohibited action-class sets, and \(C\) is the control envelope. The issued Profile is
\[
P(D) = \{\, S(D),\mspace{6mu} E,\mspace{6mu} F,\mspace{6mu} V,\mspace{6mu} I\,\}
\]
where \(E\) is the evidence record with its provenance map, \(F\) is the set of floors with their status, \(V\) is the validity function (a validity window together with the specified criteria or triggers that would deprecate the Profile's trust status\footnote{Note, these two conditions -- the validity window and the invalidation trigger -- could be treated separately as V+ and V-. The need or value of such a separation depends on the agent deployment.}), and \(I\) identifies the issuer, subject, method, schema version, and signatures. The notation makes explicit that a Profile is a structured claim with evidence and conditions, not a scalar score.

The claim is \emph{falsifiable}: each component names a value on a standardized axis, and each value determines tests that can fail---an autonomy-tier claim fails if the approval logs show reflexive acceptance; a domain claim fails if the agent proceeds after leaving its domain; an action claim fails if entitlements permit classes outside \(A^{+}\); an envelope claim fails if a named control is documented but absent from the running configuration.

The claim is \emph{bounded in time}: the validity window is for a defined period and keyed to material change, whichever is sooner, and the criteria or triggers that would void the Profile are stated (``invalidation list''), which require partial retesting.

The claim \emph{expires on boundary crossing}: it ceases to hold if the agent operates outside any element of \(S(D)\), which gives institutions notice of exactly what the claim covers and an incentive to engineer the deployment to stay within it.

Finally, the axes are \emph{jointly determining}: \(S(D)\) fixes the test environment, the evaluator suite, the failure thresholds, and the wording of the claim (Section~\protect\hyperlink{joint-effect-of-the-axes}{4.5}).

\textbf{Figure 1}. The four-axis architecture

\includegraphics[width=6.3in,height=3.79167in]{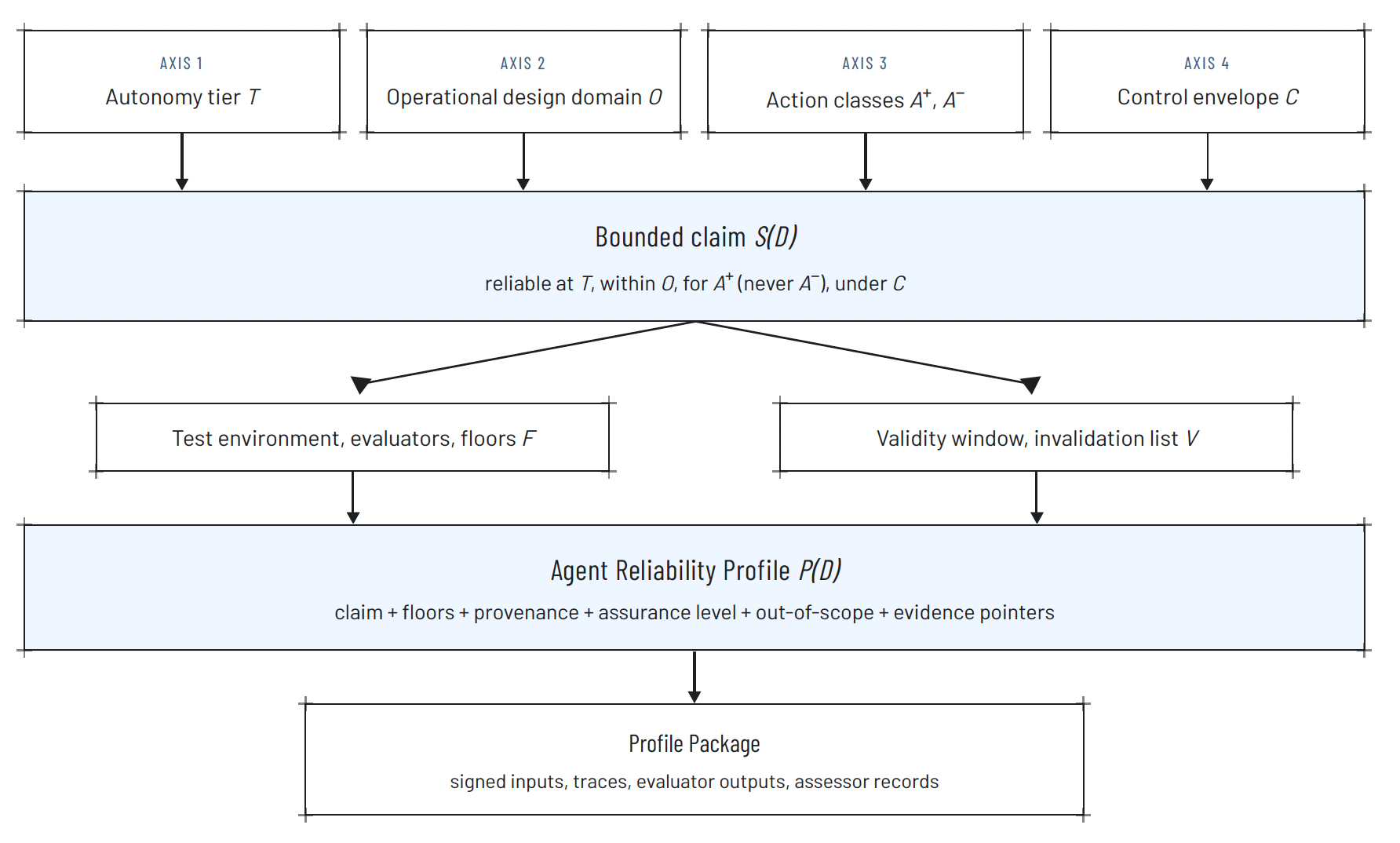}

The axes jointly fix the bounded claim, which determines the tests and the validity conditions. The issued Profile records the claim with its evidence, backed by the Profile Package.

\hypertarget{the-four-axes}{%
\section{4 The four axes}\label{the-four-axes}}

The axes correspond to properties that should be familiar to institutions deploying agents: how much a system does on its own, where it is allowed to work, what it is allowed to do, and what contains it. Our contribution is standardization---fixing the vocabulary so that two institutions describing the same deployment produce the same description.

\hypertarget{axis-1-autonomy-tier}{%
\subsection{4.1 Axis 1 --- Autonomy tier}\label{axis-1-autonomy-tier}}

We propose five tiers of autonomy, T0 to T4, and a residual status, U (unprofiled), adapted from automated driving and the convergent structure of SAE J3016 {[}10{]}, the EASA AI Concept Paper {[}11{]}, and related engineering taxonomies. Each tier is defined by two things: what the agent does, and who or what provides the fallback when the system cannot continue.

\hypertarget{tab1}{}\textbf{Table 1}. The five autonomy tiers and the residual status U

\begingroup\footnotesize
\begin{longtable}[]{@{}
  >{\raggedright\arraybackslash}p{(\columnwidth - 6\tabcolsep) * \real{0.0799}}
  >{\raggedright\arraybackslash}p{(\columnwidth - 6\tabcolsep) * \real{0.3400}}
  >{\raggedright\arraybackslash}p{(\columnwidth - 6\tabcolsep) * \real{0.2600}}
  >{\raggedright\arraybackslash}p{(\columnwidth - 6\tabcolsep) * \real{0.3200}}@{}}
\toprule\noalign{}
\begin{minipage}[t]{\linewidth}\raggedright
\textbf{Tier}
\end{minipage} & \begin{minipage}[t]{\linewidth}\raggedright
\textbf{What the agent does}
\end{minipage} & \begin{minipage}[t]{\linewidth}\raggedright
\textbf{Fallback when the agent cannot continue}
\end{minipage} & \begin{minipage}[t]{\linewidth}\raggedright
\textbf{Examples}
\end{minipage} \\
\midrule\noalign{}
\begin{minipage}[t]{\linewidth}\raggedright
T0
\end{minipage} & \begin{minipage}[t]{\linewidth}\raggedright
No action capability. Read-only or suggestion-only. A person performs the operative task.
\end{minipage} & \begin{minipage}[t]{\linewidth}\raggedright
Inapplicable---the person is the operator.
\end{minipage} & \begin{minipage}[t]{\linewidth}\raggedright
A research tool that drafts an earnings-call summary; the analyst publishes under their own name
\end{minipage} \\
\begin{minipage}[t]{\linewidth}\raggedright
T1
\end{minipage} & \begin{minipage}[t]{\linewidth}\raggedright
Assists individual actions. A person invokes each one explicitly. No chaining across a multi-step task.
\end{minipage} & \begin{minipage}[t]{\linewidth}\raggedright
The person, continuously and by construction.
\end{minipage} & \begin{minipage}[t]{\linewidth}\raggedright
A relationship manager asks the tool to draft one client email explaining a fee change, edits it, and sends it themselves
\end{minipage} \\
\begin{minipage}[t]{\linewidth}\raggedright
T2
\end{minipage} & \begin{minipage}[t]{\linewidth}\raggedright
Executes a multi-step task inside a strictly bounded envelope. A person approves every material action and is shown the specific resource that will be affected.
\end{minipage} & \begin{minipage}[t]{\linewidth}\raggedright
The person, continuously.
\end{minipage} & \begin{minipage}[t]{\linewidth}\raggedright
A loan-origination agent that pulls credit, verifies income, location, credit history, and populates the application, but pauses before each system write
\end{minipage} \\
\begin{minipage}[t]{\linewidth}\raggedright
T3
\end{minipage} & \begin{minipage}[t]{\linewidth}\raggedright
Handles the whole task inside its envelope. Requests help on uncertainty or on defined hazard triggers.
\end{minipage} & \begin{minipage}[t]{\linewidth}\raggedright
The person, on the system's request, after a handoff. The handoff is where the risk resides.
\end{minipage} & \begin{minipage}[t]{\linewidth}\raggedright
A trade-surveillance agent that closes low-scoring alerts with documented rationale, and escalates any alert involving a restricted security, or an unknown pattern
\end{minipage} \\
\begin{minipage}[t]{\linewidth}\raggedright
T4
\end{minipage} & \begin{minipage}[t]{\linewidth}\raggedright
Operates inside a bounded envelope with no contemporaneous approval. Out-of-envelope actions are blocked in software.
\end{minipage} & \begin{minipage}[t]{\linewidth}\raggedright
The system, into a defined and tested safe state, to the extent possible. A person is notified afterwards.
\end{minipage} & \begin{minipage}[t]{\linewidth}\raggedright
Real-time card-fraud decisioning: the model declines or approves within milliseconds, but can only act within pre-set rules
\end{minipage} \\
\begin{minipage}[t]{\linewidth}\raggedright
U
\end{minipage} & \begin{minipage}[t]{\linewidth}\raggedright
Unprofiled. A status, not an autonomy mode: no reliability claim is available. The envelope, the fallback, or the action boundaries are unspecified.
\end{minipage} & \begin{minipage}[t]{\linewidth}\raggedright
Undefined. U is a finding requiring remediation, suspension of agent workflows, or documented high risk acceptance.
\end{minipage} & \begin{minipage}[t]{\linewidth}\raggedright
\end{minipage} \\
\midrule\noalign{}
\endhead
\bottomrule\noalign{}
\endlastfoot
\end{longtable}
\endgroup

Each tier is defined by the agent's role in the task and by the fallback that engages when the system cannot continue. Three properties of this scale warrant emphasis.

\emph{The tier attaches to the deployment.} Following Section~\protect\hyperlink{the-deployment-as-the-unit-of-analysis}{2.1}, vendor certifications and vendor-asserted capability levels are insufficient on their own to classify a deployment. The institution's configuration choices are part of the classification.

\emph{The tier is defined by evidence.} The configuration of an agentic workflow supplies the initial evidence, and observed behavior confirms whether the asserted tier operates in practice as intended. For instance, consider the T2 designation, which involves human approval of every material action. A deployment reaches T2 when the \emph{approval log} shows that human approvals are substantive -- not based on the configuration requiring human approval. In other words, to reach T2 reviewers should be taking time over approvals, sometimes editing what is proposed, sometimes rejecting it. Where the logs show consistently reflexive acceptance, the deployment is classified according to its effective operating posture, which is T3, \emph{notwithstanding the software's configuration}. Observable indicators for this determination---time-to-approval distributions, edit rates, rejection rates, named-resource display---together with a habituation-testing protocol in which benign, materially incorrect proposed actions are periodically injected and intercepted to measure reviewer catch rates, may be important to consider{[}3{]}.

\emph{Behavior at the tier boundaries warrants the most attention.} The T2/T3 boundary is behavioral. The human-factors literature on automation, from Bainbridge's analysis of the supervisory role {[}29{]} through experimental work on automation complacency and bias {[}30{]} and the empirical record on takeover-request performance in driving autonomy, documents that a human monitoring a competent automated system degrades as a fallback: the supervisor at T3 tends to become slow, disengaged, and prone to reflexive approval. Architectural countermeasures in such a scenario may include forced engagement tied to specific hazard and regulatory triggers, and may be needed to sustain a T2 designation over time. At the T3/T4 boundary, the operative question shifts from whether the person can intervene effectively to whether the system's safe-state design bounds the potential damage. T4 evidence should detail the system's comprehensive failure management capabilities, from halting dangerous actions to limiting damage to preserving crucial data to rolling back changes to full recovery {[}3,13{]}. T4 risk often presents in different dimensions from T3: a well-architected T4 deployment with tested safe-state behavior may be safer than a T3 deployment whose human fallback is not reliable.

Six cross-cutting modifiers accompany the tier and record risk dimensions: data authority (the sensitivity of what the agent may access), reversibility profile (if agent action is reversible without a heavy penalty), customer and market impact, stack composability (whether the tool surface is predefined in advance or dynamically assembled at runtime), accountability (who stands behind the model, harness, and infrastructure as a whole), and population scope (range of parties impacted by the agentic system) {[}3{]}. These modifiers are explicitly stated on the Profile alongside the tier.

\hypertarget{axis-2-operational-design-domain}{%
\subsection{4.2 Axis 2 --- Operational design domain}\label{axis-2-operational-design-domain}}

The operational design domain (ODD), inspired by its use in automotive certification {[}10{]}, is the defined context in which the system is meant to function: the use case, the data, the operational scope, the customer population, the jurisdiction, and the cost of failure (e.g., business impacts, regulatory penalties). It specifies the test environment and is the domain-specific part of any assessment.

Unlike the other three axes, the ODD is an open set. We have identified fifty-two candidate use cases in financial services, organized under 9 business lines: (1) retail and consumer banking, (2) lending and credit, (3) payments, (4) financial crime and compliance, (5) markets and capital markets, (6) wealth and asset management, (7) treasury and finance, (8) insurance, and (9) internal/corporate functions. This list, while expansive, is preliminary. We expect it to improve and grow over time. See Appendix~\protect\hyperlink{appendix-b.-the-full-option-universe-of-the-four-axes}{B}.

For a multi-agent system, the composite domain includes the union of component domains and the inter-agent channel, whose messages, delegated tasks, and credential transfers form part of the evaluated environment.

Notably, use case boundary detection resides in this axis. An agent should be able to recognize that an observation, request, resource, or state lies outside its assessed domain and then halt, refuse, or escalate as the Profile specifies. For instance, a market research agent (ODD.30) should refuse instructions to engage in margin call related actions (ODD.34).

\hypertarget{axis-3-action-class}{%
\subsection{4.3 Axis 3 --- Action class}\label{axis-3-action-class}}

The action classes an agent can and cannot invoke materially shape its risk profile. The organizing variable is reversibility. We propose twelve classes, ordered from actions with no side effects to actions that amplify the agent's own capability surface (Table~\protect\hyperlink{tab2}{2}).

\hypertarget{tab2}{}\textbf{Table 2}. The twelve action classes, graded by reversibility and consequence.

\begingroup\footnotesize
\begin{longtable}[]{@{}
  >{\raggedright\arraybackslash}p{(\columnwidth - 6\tabcolsep) * \real{0.0799}}
  >{\raggedright\arraybackslash}p{(\columnwidth - 6\tabcolsep) * \real{0.2400}}
  >{\raggedright\arraybackslash}p{(\columnwidth - 6\tabcolsep) * \real{0.4300}}
  >{\raggedright\arraybackslash}p{(\columnwidth - 6\tabcolsep) * \real{0.2501}}@{}}
\toprule\noalign{}
\begin{minipage}[t]{\linewidth}\raggedright
\end{minipage} & \begin{minipage}[t]{\linewidth}\raggedright
\textbf{Class}
\end{minipage} & \begin{minipage}[t]{\linewidth}\raggedright
\textbf{Examples}
\end{minipage} & \begin{minipage}[t]{\linewidth}\raggedright
Reversibility
\end{minipage} \\
\midrule\noalign{}
\begin{minipage}[t]{\linewidth}\raggedright
A1
\end{minipage} & \begin{minipage}[t]{\linewidth}\raggedright
Read / retrieve
\end{minipage} & \begin{minipage}[t]{\linewidth}\raggedright
Query transactions, read case files, search policy
\end{minipage} & \begin{minipage}[t]{\linewidth}\raggedright
n/a\footnote{This ignores the risk of data leakage or theft due from cyber vulnerabilities, which are out of scope for the Profile.}
\end{minipage} \\
\begin{minipage}[t]{\linewidth}\raggedright
A2
\end{minipage} & \begin{minipage}[t]{\linewidth}\raggedright
Analyse / compute
\end{minipage} & \begin{minipage}[t]{\linewidth}\raggedright
Score, simulate, reconcile in memory
\end{minipage} & \begin{minipage}[t]{\linewidth}\raggedright
n/a
\end{minipage} \\
\begin{minipage}[t]{\linewidth}\raggedright
A3
\end{minipage} & \begin{minipage}[t]{\linewidth}\raggedright
Draft for a person
\end{minipage} & \begin{minipage}[t]{\linewidth}\raggedright
Narratives, code, emails, memos, draft filings
\end{minipage} & \begin{minipage}[t]{\linewidth}\raggedright
Fully reversible
\end{minipage} \\
\begin{minipage}[t]{\linewidth}\raggedright
A4
\end{minipage} & \begin{minipage}[t]{\linewidth}\raggedright
Recommend / route
\end{minipage} & \begin{minipage}[t]{\linewidth}\raggedright
Disposition recommendations, escalations, prioritisation
\end{minipage} & \begin{minipage}[t]{\linewidth}\raggedright
Fully reversible
\end{minipage} \\
\begin{minipage}[t]{\linewidth}\raggedright
A5
\end{minipage} & \begin{minipage}[t]{\linewidth}\raggedright
Write internal
\end{minipage} & \begin{minipage}[t]{\linewidth}\raggedright
Case notes, tickets, opening a pull request, CRM updates
\end{minipage} & \begin{minipage}[t]{\linewidth}\raggedright
Reversible
\end{minipage} \\
\begin{minipage}[t]{\linewidth}\raggedright
A6
\end{minipage} & \begin{minipage}[t]{\linewidth}\raggedright
Submit / commit internal
\end{minipage} & \begin{minipage}[t]{\linewidth}\raggedright
Merge code, close a case,
\end{minipage} & \begin{minipage}[t]{\linewidth}\raggedright
Semi-reversible
\end{minipage} \\
\begin{minipage}[t]{\linewidth}\raggedright
A7
\end{minipage} & \begin{minipage}[t]{\linewidth}\raggedright
External communication
\end{minipage} & \begin{minipage}[t]{\linewidth}\raggedright
Notify a customer, contact a third party, publish
\end{minipage} & \begin{minipage}[t]{\linewidth}\raggedright
Hard to reverse
\end{minipage} \\
\begin{minipage}[t]{\linewidth}\raggedright
A8
\end{minipage} & \begin{minipage}[t]{\linewidth}\raggedright
Transact---bounded
\end{minipage} & \begin{minipage}[t]{\linewidth}\raggedright
Place a hold, block a card, trade within a cancel window
\end{minipage} & \begin{minipage}[t]{\linewidth}\raggedright
Partially reversible
\end{minipage} \\
\begin{minipage}[t]{\linewidth}\raggedright
A9
\end{minipage} & \begin{minipage}[t]{\linewidth}\raggedright
Move funds
\end{minipage} & \begin{minipage}[t]{\linewidth}\raggedright
Wire release, payment execution, settlement
\end{minipage} & \begin{minipage}[t]{\linewidth}\raggedright
Irreversible
\end{minipage} \\
\begin{minipage}[t]{\linewidth}\raggedright
A10
\end{minipage} & \begin{minipage}[t]{\linewidth}\raggedright
Regulatory filing
\end{minipage} & \begin{minipage}[t]{\linewidth}\raggedright
File a suspicious-activity or currency report, submit returns, attest
\end{minipage} & \begin{minipage}[t]{\linewidth}\raggedright
Irreversible, with legal effect
\end{minipage} \\
\begin{minipage}[t]{\linewidth}\raggedright
A11
\end{minipage} & \begin{minipage}[t]{\linewidth}\raggedright
Access and infrastructure change
\end{minipage} & \begin{minipage}[t]{\linewidth}\raggedright
Alter permissions, deploy to production, change branch protection
\end{minipage} & \begin{minipage}[t]{\linewidth}\raggedright
Irreversible and risk-amplifying
\end{minipage} \\
\begin{minipage}[t]{\linewidth}\raggedright
A12
\end{minipage} & \begin{minipage}[t]{\linewidth}\raggedright
Delegate / spawn
\end{minipage} & \begin{minipage}[t]{\linewidth}\raggedright
Create sub-agents, grant credentials to another agent
\end{minipage} & \begin{minipage}[t]{\linewidth}\raggedright
Amplifying
\end{minipage} \\
\midrule\noalign{}
\endhead
\bottomrule\noalign{}
\endlastfoot
\end{longtable}
\endgroup

On the action axis considered alone, a fully autonomous read-only agent is a lower-risk object than a copilot that drafts a wire transfer for one-click approval. The data-authority modifier of Section~\protect\hyperlink{axis-1-autonomy-tier}{4.1} can raise the risk of a read-only deployment materially, which is why the modifiers travel with the tier.

This axis has two distinctive features. First, a deployment's value on the axis is a pair of sets: a permitted set \(A^{+}\) and a prohibited set \(A^{-}\), both explicitly stated. Second, the axis maps directly onto the identity, entitlement, and change control systems an institution already runs, which makes it examinable through access boundaries, application permissions, workflow configuration, and observed tool calls.

Multi-agent systems complicate the axis. A system composed of multiple subagents can execute composite actions that no single subagent holds---e.g., one agent retrieves data, another constructs a transaction, a third submits it. The composite Profile records the emergent class and the control points across the chain, and testing addresses the system as a whole.

\hypertarget{axis-4-control-envelope}{%
\subsection{4.4 Axis 4 --- Control envelope}\label{axis-4-control-envelope}}

The control envelope is the layered containment around the agent. It is what makes a given autonomy tier acceptable within a given domain. The layers are tested as a whole, because controls at different layers catch different failures and none is sufficient alone:

\begin{itemize}
\item
  \begin{quote}
  \emph{Model layer.} Writes out the plan of what the agent intends to do before doing anything, surfaces uncertainty, flags irreversible operations, grounds claims in retrieved evidence, and refuses out-of-scope work.
  \end{quote}
\item
  \begin{quote}
  \emph{Tool layer.} Least-privilege credentials, explicit permitted and prohibited action sets, read-only scoping of sensitive data, transaction and velocity caps, time-boxed credentials, and the absence of dangerous capabilities from the tool surface.
  \end{quote}
\item
  \begin{quote}
  \emph{Harness layer.} Human approval gates at the action, checkpoint, or batch level; immutable logging of every tool call; provenance capture preserving the evidence behind a decision; interruption hooks; automatic routing of high-risk work to senior review; and session budgets.
  \end{quote}
\item
  \begin{quote}
  \emph{Environment layer.} Sandboxing and segregation from production, network egress policy, hard execution boundaries, rollback on any committed change, encryption, retention and access logging, and immutable run records.
  \end{quote}
\item
  \begin{quote}
  \emph{Inter-agent layer}, for multi-agent systems only. Trust boundaries between agents, message validation, delegation permissions, credential isolation, and cascade limits with caps on the scope of propagated failure.
  \end{quote}
\item
  \begin{quote}
  \emph{Control-plane layer}, for platform-instantiated systems only. Registration, ownership, and versioning of agents, skills, and tools; required evaluation before publication; human review of agent-proposed changes; staged rollout, rollback, and deprecation; and change propagation with downstream Profile invalidation (Section \protect\hyperlink{reliability-as-a-bounded-engineering-claim}{2.2}).
  \end{quote}
\end{itemize}

The full catalogue lists fifty-five named controls across the six layers (Appendix~\protect\hyperlink{appendix-b.-the-full-option-universe-of-the-four-axes}{B}) incorporating well-established frameworks like OWASP and NIST. An agent's Profile records which controls are present, how each is configured, the source of the assertion, and the validation method and result. Documentation alone can support a Level~1 assertion, while higher levels require detailed evidence from inspections, evaluations, or tests.

On this axis the Profile is descriptive: it records the envelope an institution has chosen and tests whether the envelope is real, leaving the choice of envelope to the institution and its regulators.\footnote{The Monetary Authority of Singapore's SAFR framework {[}16{]} takes a more prescriptive approach to runtime controls; see Section~\protect\hyperlink{deployment-disclosure-and-assurance}{9.3}.}

\hypertarget{joint-effect-of-the-axes}{%
\subsection{4.5 Joint effect of the axes}\label{joint-effect-of-the-axes}}

The four axes operate as a single test specification. The autonomy tier dictates the response to failure, covering human handoffs, required approvals, and guaranteed fail-safe modes. The ODD establishes the verifiable operating context, or ``ground truth'' for the system. The action sets define the possible actions the system can perform, the resulting consequences of those actions, and the necessary recovery steps. And the control envelope determines which safety protocols and limitations must be observed operating during the test.

By combining these four axes, we create an actionable test plan that is complete: the test scenario, the permitted and forbidden actions, the required documentation, the passing criteria, and the precise boundaries of the system's claimed performance. Together, they create a comprehensive, auditable blueprint. The blueprint specifies the entire test scenario, the rules for system operations, the required safety checks, the minimum performance standards, and the exact scope of the final safety claim.

\hypertarget{the-agent-reliability-profile-artifact}{%
\section{5 The Agent Reliability Profile artifact}\label{the-agent-reliability-profile-artifact}}

The output of the architecture is a single artifact: the Agent Reliability Profile. The Profile is the concise issued summary for one agentic deployment.

\hypertarget{profile-structure}{%
\subsection{5.1 Profile structure}\label{profile-structure}}

Table~\protect\hyperlink{tab3}{3} defines the minimum content of a conforming Profile. To include organizational level policies and requirements, additional components would likely need to be considered and added.

\hypertarget{tab3}{}\textbf{Table 3}. Minimum content of a conforming Agent Reliability Profile

\begingroup\footnotesize
\begin{longtable}[]{@{}
  >{\raggedright\arraybackslash}p{(\columnwidth - 4\tabcolsep) * \real{0.2000}}
  >{\raggedright\arraybackslash}p{(\columnwidth - 4\tabcolsep) * \real{0.5000}}
  >{\raggedright\arraybackslash}p{(\columnwidth - 4\tabcolsep) * \real{0.3000}}@{}}
\toprule\noalign{}
\begin{minipage}[t]{\linewidth}\raggedright
\textbf{Profile section}
\end{minipage} & \begin{minipage}[t]{\linewidth}\raggedright
\textbf{Required content}
\end{minipage} & \begin{minipage}[t]{\linewidth}\raggedright
\textbf{Purpose}
\end{minipage} \\
\midrule\noalign{}
\begin{minipage}[t]{\linewidth}\raggedright
Identity and issuance
\end{minipage} & \begin{minipage}[t]{\linewidth}\raggedright
Profile identifier; deployment identity; name of accountable owner; name of issuer; issue time; validity interval; schema and versions; signatures of the owner and issuer; parent Platform Baseline Profile and version, where applicable
\end{minipage} & \begin{minipage}[t]{\linewidth}\raggedright
Binds the claim to a deployment, owner, issuer, and method
\end{minipage} \\
\begin{minipage}[t]{\linewidth}\raggedright
Claim boundaries
\end{minipage} & \begin{minipage}[t]{\linewidth}\raggedright
Autonomy tier; operational design domain; permitted and prohibited action-class sets; control envelope
\end{minipage} & \begin{minipage}[t]{\linewidth}\raggedright
States the assessed scope
\end{minipage} \\
\begin{minipage}[t]{\linewidth}\raggedright
Assurance statement
\end{minipage} & \begin{minipage}[t]{\linewidth}\raggedright
Production assurance level (Level 1, 2, or 3) for each axis and the headline level; benchmark comparability flag, with its environment-fidelity statement where set; assessor role; test-environment
\end{minipage} & \begin{minipage}[t]{\linewidth}\raggedright
Identifies the basis and independence of the evidence
\end{minipage} \\
\begin{minipage}[t]{\linewidth}\raggedright
Floors
\end{minipage} & \begin{minipage}[t]{\linewidth}\raggedright
Required pass-or-fail conditions; status; finding references
\end{minipage} & \begin{minipage}[t]{\linewidth}\raggedright
Prevents aggregate performance from masking critical failure
\end{minipage} \\
\begin{minipage}[t]{\linewidth}\raggedright
Results
\end{minipage} & \begin{minipage}[t]{\linewidth}\raggedright
Evaluator-level results at assurance level 2 or 3; benchmark result where the benchmark comparability flag is set
\end{minipage} & \begin{minipage}[t]{\linewidth}\raggedright
Reports supported findings without implying additional assurance
\end{minipage} \\
\begin{minipage}[t]{\linewidth}\raggedright
Provenance map
\end{minipage} & \begin{minipage}[t]{\linewidth}\raggedright
Field-level source type: measured, configuration-derived, document-derived, owner-attested, or absent
\end{minipage} & \begin{minipage}[t]{\linewidth}\raggedright
Makes evidence quality visible
\end{minipage} \\
\begin{minipage}[t]{\linewidth}\raggedright
Out-of-scope statement
\end{minipage} & \begin{minipage}[t]{\linewidth}\raggedright
Domains, actions, controls, and users / adopters excluded from the claim
\end{minipage} & \begin{minipage}[t]{\linewidth}\raggedright
Defines the external boundary of the claim
\end{minipage} \\
\begin{minipage}[t]{\linewidth}\raggedright
Invalidation list
\end{minipage} & \begin{minipage}[t]{\linewidth}\raggedright
Changes that void the Profile, or might require partial retesting,
\end{minipage} & \begin{minipage}[t]{\linewidth}\raggedright
Supports change-sensitive validity
\end{minipage} \\
\begin{minipage}[t]{\linewidth}\raggedright
Evidence pointers
\end{minipage} & \begin{minipage}[t]{\linewidth}\raggedright
Content hashes and stable identifiers for inputs, evaluator outputs and action traces, and assessor reports
\end{minipage} & \begin{minipage}[t]{\linewidth}\raggedright
Connects the summary to the Profile Package
\end{minipage} \\
\begin{minipage}[t]{\linewidth}\raggedright
Status and limitations
\end{minipage} & \begin{minipage}[t]{\linewidth}\raggedright
Open findings, exceptions, risk acceptances, and interpretation limits
\end{minipage} & \begin{minipage}[t]{\linewidth}\raggedright
Preserves decision context
\end{minipage} \\
\midrule\noalign{}
\endhead
\bottomrule\noalign{}
\endlastfoot
\end{longtable}
\endgroup

Profiles are explicitly designed to not over-state any claims. A Profile states only conclusions its assurance level supports. Comparative grades appear only on Benchmarked Profiles, where the shared environment, scenario policy, and scoring method support comparison across deployments. Every other Profile states plainly that its results are specific to its own environment.

\hypertarget{provenance-and-the-profile-package}{%
\subsection{5.2 Provenance and the Profile Package}\label{provenance-and-the-profile-package}}

Every material Profile field carries provenance: measured, checked against deployed configuration, extracted from a signed institutional document, attested by a named owner, or absent. Conflicts are made visible. For instance, if a policy specifies read-only access but the deployed credential holds write permissions, the Level~1 Profile records the conflict, while the Level~2 Profile tests the effective boundary.

The Profile Package includes the broader set of information needed to support a Profile. It includes, as applicable: the signed source manifest and inputs; the deployment bill of materials and environment manifest; the tool catalog, permission map, and control configuration; the domain specification and scenario manifest; evaluator definitions, versions, and thresholds; the immutable execution trace; floor findings and evaluator outputs; replay material (where disclosure is safe); the independent assessor's report; benchmark environment and hidden-set version identifiers; and the cryptographic hashes and signatures connecting each component.

\hypertarget{validity-and-change-control}{%
\subsection{5.3 Validity and change control}\label{validity-and-change-control}}

The Profile has a validity interval, i.e., a specified period of time for which the profile is applicable and a machine-readable invalidation list, i.e., specific conditions or triggers that would make the profile invalid. Changes that ordinarily void a Profile include expansion to a new ODD, addition of a higher-consequence action class, removal of a critical control, new access such a revised owner , a changed multi-agent topology with new delegation paths, and a model or harness change that can alter tested behavior.\footnote{One could also flip the presumption and automatically revoke the Profile every N days unless the Profile is re-asserted or re-validated based on fresh information.}

In the case of platform-instantiated agents, a material change to the platform substrate voids the Platform Baseline Profile and, with it, every anchored Profile. Post-deployment monitoring determines whether the issued conditions continue to hold; NIST AI 800-4 explains some of the difficulty of drift detection, distributed logging, and live monitoring for deployed AI systems {[}21{]}, and the Profile surfaces these difficulties explicitly through its monitoring plan, and reissuance requirements. A point-in-time evaluation is treated as exactly that, and the residual risk between observations is acknowledged in Section~\protect\hyperlink{limitations-and-open-questions}{10}.

Worked examples of the bounded claim, for a coding agent and for a financial-crime alert-triage agent, appear in Appendix~\protect\hyperlink{appendix-a.-the-bounded-claim}{A}.

\hypertarget{graduated-assurance-three-levels-a-comparability-flag-and-their-tools}{%
\section{6 Graduated assurance: three levels, a comparability flag, and their tools}\label{graduated-assurance-three-levels-a-comparability-flag-and-their-tools}}

Not all agents need the same level of assurance as to their reliability. Higher risk agents and those impacting a wide range of stakeholders may warrant greater assurance as to the validity of the claims in their Profiles. The proper level of assurance will depend on the use case, the firm, and the deployment context.

We propose three levels of production assurance---Level 1, Level 2, and Level 3---and a separate benchmark comparability flag. Levels 1 through 3 form a single ordered scale of evidence and independence about the deployment in its own environment. The benchmark flag is a different measurement, trading production fidelity for cross-institution comparability. This may carry higher assurance value, but does not in all cases, which is why we separate it. The decision of what level of assurance to seek may be made by the deploying firm or it might be required by a standards or regulatory body. The Profile keeps the same shape at every level---the axes, components, and definitions are standardized---while the rigor and independence of the evidence adjust.

\textbf{Table 4}. The three production assurance levels and the benchmark comparability flag

\begingroup\footnotesize
\begin{longtable}[]{@{}
  >{\raggedright\arraybackslash}p{(\columnwidth - 8\tabcolsep) * \real{0.1600}}
  >{\raggedright\arraybackslash}p{(\columnwidth - 8\tabcolsep) * \real{0.2000}}
  >{\raggedright\arraybackslash}p{(\columnwidth - 8\tabcolsep) * \real{0.2000}}
  >{\raggedright\arraybackslash}p{(\columnwidth - 8\tabcolsep) * \real{0.2200}}
  >{\raggedright\arraybackslash}p{(\columnwidth - 8\tabcolsep) * \real{0.2200}}@{}}
\toprule\noalign{}
\begin{minipage}[t]{\linewidth}\raggedright
\end{minipage} & \begin{minipage}[t]{\linewidth}\raggedright
\textbf{Level 1}

Asserted Profile
\end{minipage} & \begin{minipage}[t]{\linewidth}\raggedright
\textbf{Level 2}

Validated Profile
\end{minipage} & \begin{minipage}[t]{\linewidth}\raggedright
\textbf{Level 3}

Verified Profile
\end{minipage} & \begin{minipage}[t]{\linewidth}\raggedright
Comparability flag

Benchmarked Profile
\end{minipage} \\
\midrule\noalign{}
\begin{minipage}[t]{\linewidth}\raggedright
What happens
\end{minipage} & \begin{minipage}[t]{\linewidth}\raggedright
The institution's own documents and settings are read and compiled into a Profile
\end{minipage} & \begin{minipage}[t]{\linewidth}\raggedright
The institution runs tests against its own agent
\end{minipage} & \begin{minipage}[t]{\linewidth}\raggedright
A qualified independent assessor runs the tests
\end{minipage} & \begin{minipage}[t]{\linewidth}\raggedright
An independent assessor runs the agent in a shared environment
\end{minipage} \\
\begin{minipage}[t]{\linewidth}\raggedright
Who stands behind it
\end{minipage} & \begin{minipage}[t]{\linewidth}\raggedright
The institution
\end{minipage} & \begin{minipage}[t]{\linewidth}\raggedright
The institution
\end{minipage} & \begin{minipage}[t]{\linewidth}\raggedright
An independent assurance firm
\end{minipage} & \begin{minipage}[t]{\linewidth}\raggedright
An independent assessor
\end{minipage} \\
\begin{minipage}[t]{\linewidth}\raggedright
Where it runs
\end{minipage} & \begin{minipage}[t]{\linewidth}\raggedright
Nothing is executed
\end{minipage} & \begin{minipage}[t]{\linewidth}\raggedright
The institution's own environment
\end{minipage} & \begin{minipage}[t]{\linewidth}\raggedright
The institution's own environment
\end{minipage} & \begin{minipage}[t]{\linewidth}\raggedright
A shared reference environment with a hidden, rotating scenario set
\end{minipage} \\
\begin{minipage}[t]{\linewidth}\raggedright
What the Profile says
\end{minipage} & \begin{minipage}[t]{\linewidth}\raggedright
Self-reported
\end{minipage} & \begin{minipage}[t]{\linewidth}\raggedright
Self-reported, self-tested
\end{minipage} & \begin{minipage}[t]{\linewidth}\raggedright
Self-reported, independently tested
\end{minipage} & \begin{minipage}[t]{\linewidth}\raggedright
In addition to its production assurance level: independently benchmarked
\end{minipage} \\
\begin{minipage}[t]{\linewidth}\raggedright
Minimum bar to claim it
\end{minipage} & \begin{minipage}[t]{\linewidth}\raggedright
Inputs attested by a named owner
\end{minipage} & \begin{minipage}[t]{\linewidth}\raggedright
Reproducible generation, a content hash, signed inputs
\end{minipage} & \begin{minipage}[t]{\linewidth}\raggedright
The Level 2 bar, plus a qualified independent assessor and its own signed report
\end{minipage} & \begin{minipage}[t]{\linewidth}\raggedright
Accredited assessor, named environment version, named hidden-set version
\end{minipage} \\
\begin{minipage}[t]{\linewidth}\raggedright
What it may claim
\end{minipage} & \begin{minipage}[t]{\linewidth}\raggedright
This is what we believe we have deployed
\end{minipage} & \begin{minipage}[t]{\linewidth}\raggedright
No critical control failure under the conditions we tested
\end{minipage} & \begin{minipage}[t]{\linewidth}\raggedright
The same, tested by someone other than us
\end{minipage} & \begin{minipage}[t]{\linewidth}\raggedright
A result comparable with other institutions under the reference conditions
\end{minipage} \\
\begin{minipage}[t]{\linewidth}\raggedright
What it may claim beyond that
\end{minipage} & \begin{minipage}[t]{\linewidth}\raggedright
Nothing tested
\end{minipage} & \begin{minipage}[t]{\linewidth}\raggedright
No independence
\end{minipage} & \begin{minipage}[t]{\linewidth}\raggedright
No comparability with any other system
\end{minipage} & \begin{minipage}[t]{\linewidth}\raggedright
No additional assurance about production behavior
\end{minipage} \\
\begin{minipage}[t]{\linewidth}\raggedright
Does data leave the institution
\end{minipage} & \begin{minipage}[t]{\linewidth}\raggedright
No
\end{minipage} & \begin{minipage}[t]{\linewidth}\raggedright
No
\end{minipage} & \begin{minipage}[t]{\linewidth}\raggedright
Either the evidence travels or the assessor works on site
\end{minipage} & \begin{minipage}[t]{\linewidth}\raggedright
Sometimes, depending on the benchmarking infrastructure and the benchmarked components
\end{minipage} \\
\midrule\noalign{}
\endhead
\bottomrule\noalign{}
\endlastfoot
\end{longtable}
\endgroup

The Profile's structure is constant; the rigor and independence of the evidence adjust across the levels, and the flag is reported separately.

\textbf{Figure 2}. The assurance ladder

\includegraphics[width=6.3in,height=2.5in]{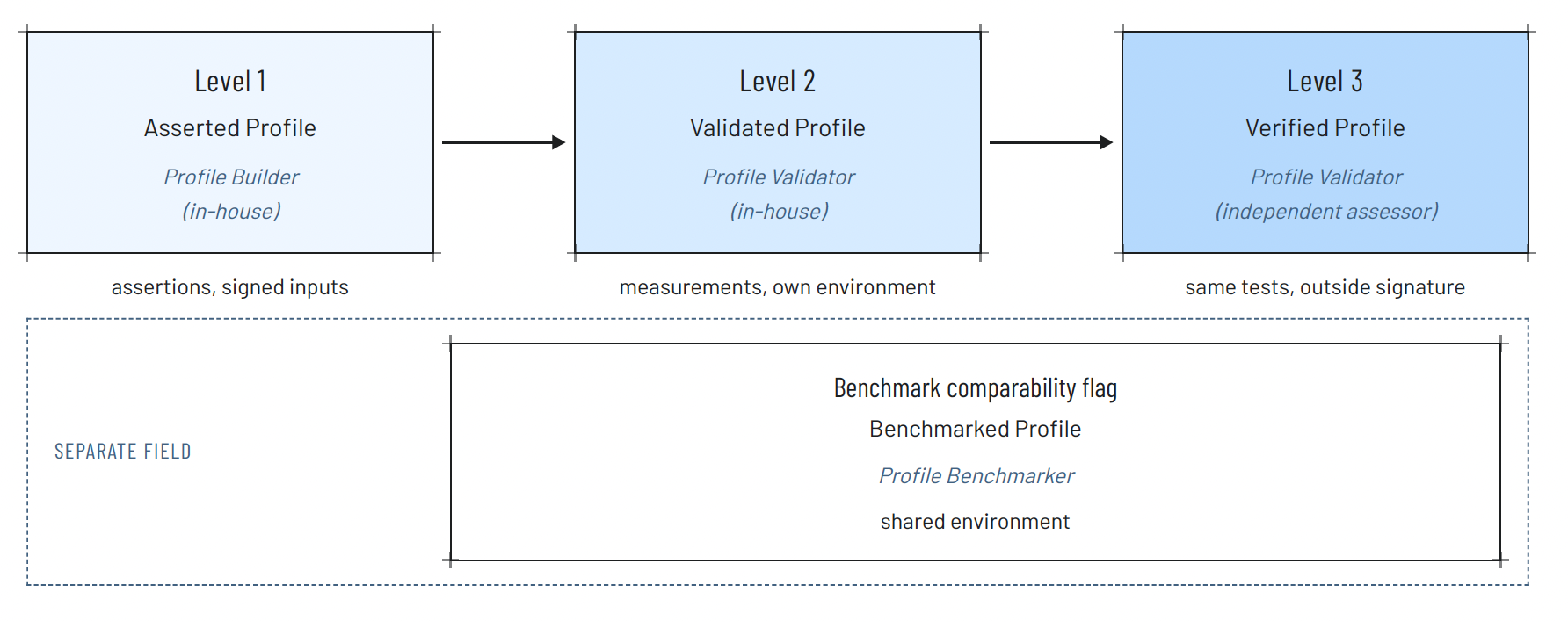}

While the assurance gains of moving from Asserted to Validated to Verified Profiles are intuitive and their production is tractable today, Benchmarked Profiles are situated differently. From a technical perspective, more work needs to be done to define the proper scope for systems-under-test (SUTs) for agentic deployments consistent with robust benchmarking science. From an ecosystem perspective, the comparability feature of Benchmarked Profiles has the potential to enable assurance at scale in ways that Level 1-3 Profiles might find challenging. For instance, banks lacking the resources to evaluate agents they wish to deploy might find it more possible to achieve assurance if relevant benchmarks are developed. Both the technical and ecosystem aspects of reliability benchmarking warrant further research and development.

\hypertarget{the-tools}{%
\subsection{6.1 The tools}\label{the-tools}}

Each rung of the ladder, and the comparability flag, is served by a defined tool.

The \emph{Profile Builder} produces the Level~1 Asserted Profile. It ingests a signed source manifest and parses the institution's governance documents, architecture records, tool and credential settings, and control configurations; maps fields to the four-axis schema; compares intended controls with as-built settings; identifies conflicts and unsupported fields; and drafts the bounded claim and a field-level provenance map. Nothing is executed. Deterministic parsers handle structured configuration; grounded extraction may assist with narrative documents, with citations to source locations and an explicit unsupported state. A named owner reviews the result, resolves or accepts conflicts, and issues the Asserted Profile with a signature. The Builder is designed to make Level~1 Profiles inexpensive to generate so that an institution can inventory its full agentic estate at this level.

The \emph{Profile Validator} produces the Level~2 Validated Profile. Run in-house, it executes scenarios against the deployed agent through a deployment adapter, with a layered policy broker controlling the actions exposed during evaluation. Cross-cutting evaluators measure prohibited-action attempts, tool discipline, boundary detection, escalation, trace completeness, and safe-state behavior; domain evaluators measure task-specific correctness and harm conditions. Inputs, method versions, environment identifiers, and outputs are hashed and signed to support reproduction. Assertions from Level~1 Profiles become measurements, which makes material misstatement detectable.

Level~3 introduces testing independence. The same Profile Validator, with the same conformance requirements, is operated and signed by a qualified independent assessor, producing the Verified Profile. Everything else is held constant. The independence model requires assessor qualification, conflict rules, secure operation on site or with transferred evidence, retention, and the role of a registry (see Section~\protect\hyperlink{implementation-program-and-research-agenda}{11}).

The \emph{Profile Benchmarker} sets the benchmark comparability flag, producing a Benchmarked Profile. An independent assessor runs the agent through a shared reference environment and a hidden, rotating scenario set, with the environment version, tool surface, data snapshot, scoring method, floor specification, and hidden-set version all controlled. Because the environment and scenarios are shared, results become comparable across institutions.

Terminology follows the same discipline throughout. An institution \emph{assesses} an agent and \emph{compiles} its Profile; a named owner \emph{issues} and signs it. \emph{Validating} (Level~2) and \emph{benchmarking} (the comparability flag) are the evidence-generating operations; \emph{verifying} (Level~3) is what a qualified independent assessor does; \emph{examining} is what a supervisor does to any Profile. The artifact's cryptographic signature is separately checked by the relying party's systems.

\hypertarget{mixed-axis-assurance-the-headline-level-and-the-comparability-flag}{%
\subsection{6.2 Mixed-axis assurance, the headline level, and the comparability flag}\label{mixed-axis-assurance-the-headline-level-and-the-comparability-flag}}

Production assurance can differ by axis: an institution may obtain independent verification for the axis a supervisor or counterparty is focused on while self-testing the rest. The Profile records a production assurance vector
\[
L(P) = \left( L_{T},\mspace{6mu} L_{O},\mspace{6mu} L_{A},\mspace{6mu} L_{C} \right)
\]
with one value for each axis---autonomy tier (T), ODD (O), action class (A), and control envelope (C)---each drawn from the ordered set Level 1 \textless{} Level 2 \textless{} Level 3. The headline level is the minimum:
\[
L_{headline}(P) = \min\{ L_{T},\mspace{6mu} L_{O},\mspace{6mu} L_{A},\mspace{6mu} L_{C}\}.
\]
The rule prevents strong evidence on one axis from raising the apparent assurance of a weaker axis, while the full vector remains visible. Because the three production assurance levels are totally ordered, the minimum is always defined. The benchmark comparability flag does not enter it: the flag is recorded as a separate field. A Profile benchmarked on one axis and independently verified on another has a well-defined headline level and a separately reported benchmark result.

Levels 1 and 2 should be inexpensive to produce and are accordingly most useful for lower-risk agents and estate-wide inventory. Level 3 costs more, provides elevated assurance, and suits higher-risk agents. Benchmarking is the costliest to produce and buys comparability. One asymmetry across the axes deserves note: the autonomy axis is the one on which the institution's own operational evidence remains indispensable even at the highest assurance levels, because distinguishing substantive approval from reflexive approval requires the approval logs of the actual deployment.

For platform-instantiated systems, assurance also composes vertically. Independent verification at Level 3, obtained once for the Platform Baseline Profile, covers the substrate for every anchored agent. The anchored Profiles above it can remain Level 1 artifacts generated from control-plane configuration, upgraded selectively where a deployment's risk warrants it. This composition makes profiling a large portfolio of platform-derived agents economically tractable.

\hypertarget{evaluation-methodology}{%
\section{7 Evaluation methodology}\label{evaluation-methodology}}

\hypertarget{test-plan}{%
\subsection{7.1 Test plan}\label{test-plan}}

A test plan explains how an agent's reliability claims will be checked. It identifies the situations to test, the behavior and safeguards to observe, and the evidence needed to judge the results.

The four axes of the Level 1 Profile shape the test plan. A planning tool first identifies the deployed agent and its component versions. The agent's autonomy level determines which approval, handoff, fallback, and safe-state tests apply. Its operating setting determines the data, tools, people, jurisdictions, and consequences of failure the tests must reflect. The plan then links permitted and prohibited actions to specific tool operations and checks whether required controls work. It sets the evidence needed, minimum acceptable results, number of runs, and ways to vary test conditions.

After testing, reviewers preserve the record, assess the findings, and state what the evidence supports. The plan also explains what each test measures and which likely failures it may miss. Completing a task is only one measure: tests also examine the agent's decisions, tool use, rule compliance, boundaries, response to failure, recovery, and consistency across runs {[}2,17{]}.

\hypertarget{scenario-classes-and-coverage}{%
\subsection{7.2 Scenario classes and coverage}\label{scenario-classes-and-coverage}}

Testing agents requires scenarios. A scenario gives the agent a specific task under defined conditions so evaluators can observe what it does. Subjecting agents to a range of scenarios reveals failures that routine tasks may miss. For example, a payment agent might complete a standard transfer but need to pause and seek approval when the amount exceeds its authority.

Scenario coverage and balance matter. A balanced set of scenarios tests normal work and situations that may expose failure. It includes routine and difficult tasks within the agent's intended scope, cases near or outside that scope, and attempts to induce prohibited actions. Tests also cover tool, data, or network failures; unclear or conflicting instructions; hostile inputs or hidden instructions in outside material; outdated or inconsistent data; delayed human handoffs; failures in communication or delegation between agents; and situations that require the agent to pause safely or recover. The coverage record shows which operating conditions, actions, possible consequences, evaluation methods, and safeguards were tested. It flags gaps so readers can judge what the results support and what remains unknown.

\hypertarget{repeated-runs-and-reliability-metrics}{%
\subsection{7.3 Repeated runs and reliability metrics}\label{repeated-runs-and-reliability-metrics}}

Because the models powering agents are nondeterministic, agents may respond differently to the same task. Thus, repeated tests are needed to show whether performance holds up. Additional runs may be warranted where results vary or mistakes carry greater consequences.

When doing repeated runs, the following reliability metrics are fairly standard and should be utilized: task success across runs; consistency across equivalent inputs; robustness to controlled perturbations; rule compliance (policy conformance rate); attempted and blocked prohibited actions (interception rates); boundary detection performance; and confidence accuracy, where measurable. In addition, test reports should include recurring failures, the frequency and severity of errors, recovery success and time to safety, and the completeness of each run's record. These measures are outlined by Rabanser et al. {[}2{]}.

The Profile keeps metrics disaggregated, rather than converting them into a composite score. A set of minimum acceptable results (floors) for issuance should be established prior to runs.

\hypertarget{evaluating-human-in-the-loop-approval}{%
\subsection{7.4 Evaluating human-in-the-loop approval}\label{evaluating-human-in-the-loop-approval}}

One of the most challenging claims in the Profile to evaluate is the autonomy level of an agent, especially when it is contingent on human approval, such as with human-in-the-loop arrangements (T2). If the human's approvals are pro forma, then the de facto autonomy level of the agent is higher (say, T3) than the configuration or policy says. The effectiveness of human approval is most salient.

A significant body of research shows that the effectiveness of human approval as a control varies with workload, interface design, familiarity, and time {[}29,30{]}. Accordingly, evidence for a T2 claim includes approval latency, edit rate, rejection rate, resource display, consequence display, and sampled review quality. Habituation testing measures whether reviewers detect intercepted, benign, materially incorrect proposals, with test actions blocked before consequence {[}3{]}. At T3, the evaluation measures handoff performance directly: trigger accuracy, time to acknowledgment, time to competent control, information transferred, state preservation, and actions taken during the delay.

\hypertarget{environment-fidelity-and-transfer-uncertainty}{%
\subsection{7.5 Environment fidelity and transfer uncertainty}\label{environment-fidelity-and-transfer-uncertainty}}

Levels 2 and 3 maximize deployment fidelity, because tests run in the institution's environment. However, this limits cross-firm comparability. Benchmarking achieves comparability by maximizing environmental standardization. But this introduces transfer uncertainty because the reference environment differs from production.

Each result therefore states how closely the test setting matched the real deployment: its tools, data, policies, access permissions, response times, failure conditions, population covered, and any known differences.

\hypertarget{integrity-reproducibility-and-confidentiality}{%
\subsection{7.6 Integrity, reproducibility, and confidentiality}\label{integrity-reproducibility-and-confidentiality}}

Each evaluation fixes and records the software, data formats, policies, evaluation tools, data snapshots, and random settings it uses. Repeating a test requires stable inputs; when inputs are confidential, a cryptographic commitment can help verify that they have not changed. Because hidden benchmark tasks and institutional data cannot always be shared, the Profile Package retains version identifiers and selected evidence without revealing sensitive content.

Keeping a benchmark trustworthy also takes ongoing work: reviewing transcripts, defining what agents may access or do, limiting data that leaves the test environment, protecting correct answers, testing scorers against manipulation, refreshing scenarios, and investigating unusual successes {[}19,20{]}. NIST's AI Agent Standards Initiative identifies security evaluations, agent identity, and interoperability as priorities {[}24{]}. The Profile offers a possible way to connect those elements to a claim about a specific deployment.

\hypertarget{use-in-financial-services}{%
\section{8 Use in financial services}\label{use-in-financial-services}}

\hypertarget{by-financial-institutions}{%
\subsection{8.1 By financial institutions}\label{by-financial-institutions}}

Agent Reliability Profiles can address several interrelated challenges faced by financial institutions:

\emph{Internal risk management} -- Many institutions today have difficulty answering, in consistent terms, what their agents are permitted to do and what contains them. An inventory of agentic deployments, each with a stated bounded claim supported by a Profile, would provide a board, a chief risk officer, or risk committee, with visibility into their institution's agent-related activities, in the same language used by supervisors and external parties. The risk management focus of the Profile on reliable agent functioning within boundaries would also balance and complement internal reporting on agent performance and token efficiency.

\emph{Acceptance by users and third party partners} -- To the extent an institution's agents are customer-facing or interact with external parties, such as a peer financial institution or vendor, the Profile provides a standardized way to convey the boundaries of its agents (Level 1), as well as evidence of validation (Levels 2 and 3). The standardized taxonomy lowers the cost of those parties' due diligence, enabling greater trust -- especially for agents with Level 2 or 3 Profiles -- and thus accelerated acceptance and willingness to adopt or interact with those agents.

\emph{Due diligence of third party and vendor agents} -- Financial institutions could utilize the Profile to improve and accelerate due diligence of agents from third party peers and vendors. Without a standardized Profile, financial institutions must learn the operating boundary vocabulary and controls of each agent developer, limiting the speed and scale of such engagements.

\emph{Supervisory expectations and assurance} -- The standardized taxonomy underlying the Profile can facilitate familiarity and consistency of examinations and supervisory reviews of agents deployed at supervised institutions. In the U.S., the federal banking agencies in April 2026 excluded generative and agentic AI from model risk management guidance (SR 26-2). While this has provided greater flexibility to banks, it has also created a vacuum in terms of supervisory expectations. The Profile begins to address that gap with its standardized taxonomy, focus on controls, and support of validation and third party verification. Its sponsorship by MLCommons, a non-profit, also mitigates commercial conflicts of interest related to its adoption.

\hypertarget{by-supervisors}{%
\subsection{8.2 By supervisors}\label{by-supervisors}}

The Profile can help supervisors overcome challenges with assessing and examining the safety and soundness of agents deployed by supervised institutions. Namely, it provides:

\begin{enumerate}
\def\labelenumi{\arabic{enumi}.}
\item
  \begin{quote}
  \textbf{A common vocabulary}. Currently, each financial institution utilizes its own terminology and taxonomy to capture, communicate, and monitor the risks of its agents, limiting the ability of supervisors to learn and oversee agents at scale across institutions. The Profile provides a standardized taxonomy and framework for describing the reliability of agents.
  \end{quote}
\item
  \begin{quote}
  \textbf{An examinable artifact}. The Profile is structured around testable claims that are directly relevant to the reliability of the agent. The Profile Package assembles the relevant supporting materials to substantiate those claims and any tests done against them, thus allowing supervisors to focus on outcomes rather than process.
  \end{quote}
\item
  \begin{quote}
  \textbf{Practical risk tiering}. The Profile enables users to categorize agents by riskiness along the four axes -- autonomy level, operational design domain, action class, and control envelope. Because Profiles are standardized, risk tiering can be applied across cohorts of agents, even from different institutions.
  \end{quote}
\item
  \begin{quote}
  \textbf{A basis for trend and outlier analysis}. The adoption of Profiles by institutions would allow supervisors to monitor trends in agentic deployments and to identify outliers amongst institutions. Moreover, examinations of an institution's Level 2 Profiles could provide supervisory insight into the robustness of its testing and validation functions and support broader supervisory assessments of an institution's AI-related risk management and control capabilities.
  \end{quote}
\end{enumerate}

\hypertarget{the-role-of-the-profile-in-agent-assurance}{%
\section{9 The role of the Profile in agent assurance}\label{the-role-of-the-profile-in-agent-assurance}}

Agent assurance has developed and evolved across several layers. Each layer addresses a different question and unit of analysis (Table 5). For instance, model and product evaluation asks whether a model or upstream product performs well on defined tasks. Institutional risk governance asks whether an enterprise manages AI risk through sound processes. Identity schemes ask which agent is acting and on whose authority. Runtime controls ask whether a particular agentic action should proceed at the moment it is attempted.

Taken together, these layers can bolster trust in agents. But the agents themselves exist between the two institutional layers and the two action layers described above. That deployment layer -- consisting of a specific configuration of model, harness, tools, credentials, and environment, operated by a specific institution for a specific purpose -- lacks a standardized, testable assurance mechanism.

The Agent Reliability Profile aims to fill that gap and serve as a bridge between the institutional and action layers supporting agent assurance (Table 5). Each layer answers a different question about a different unit of analysis. The Agent Reliability Profile supplies the deployment layer.

\textbf{Table 5}. Layers of agent assurance

\begingroup\footnotesize
\begin{longtable}[]{@{}
  >{\raggedright\arraybackslash}p{(\columnwidth - 12\tabcolsep) * \real{0.1364}}
  >{\raggedright\arraybackslash}p{(\columnwidth - 12\tabcolsep) * \real{0.1742}}
  >{\raggedright\arraybackslash}p{(\columnwidth - 12\tabcolsep) * \real{0.1250}}
  >{\raggedright\arraybackslash}p{(\columnwidth - 12\tabcolsep) * \real{0.1193}}
  >{\raggedright\arraybackslash}p{(\columnwidth - 12\tabcolsep) * \real{0.1307}}
  >{\raggedright\arraybackslash}p{(\columnwidth - 12\tabcolsep) * \real{0.1496}}
  >{\raggedright\arraybackslash}p{(\columnwidth - 12\tabcolsep) * \real{0.1648}}@{}}
\toprule\noalign{}
\begin{minipage}[t]{\linewidth}\raggedright
Layer
\end{minipage} & \begin{minipage}[t]{\linewidth}\raggedright
Question it answers
\end{minipage} & \begin{minipage}[t]{\linewidth}\raggedright
Unit of analysis
\end{minipage} & \begin{minipage}[t]{\linewidth}\raggedright
When it operates
\end{minipage} & \begin{minipage}[t]{\linewidth}\raggedright
Who stands behind it
\end{minipage} & \begin{minipage}[t]{\linewidth}\raggedright
Representative instruments
\end{minipage} & \begin{minipage}[t]{\linewidth}\raggedright
Relationship to the Profile
\end{minipage} \\
\midrule\noalign{}
\begin{minipage}[t]{\linewidth}\raggedright
Model and product evaluation
\end{minipage} & \begin{minipage}[t]{\linewidth}\raggedright
Is this model or agentic product capable and well-behaved on defined tasks?
\end{minipage} & \begin{minipage}[t]{\linewidth}\raggedright
Model or product
\end{minipage} & \begin{minipage}[t]{\linewidth}\raggedright
Before release and at version changes
\end{minipage} & \begin{minipage}[t]{\linewidth}\raggedright
Developer or vendor; benchmark bodies
\end{minipage} & \begin{minipage}[t]{\linewidth}\raggedright
Model and system cards {[}27{]}; product benchmarks, including the agentic product maturity ladder {[}1{]}; evaluation infrastructure {[}17,18{]}
\end{minipage} & \begin{minipage}[t]{\linewidth}\raggedright
Upstream evidence. A product result is an input to a Profile claim, never a substitute for one.
\end{minipage} \\
\begin{minipage}[t]{\linewidth}\raggedright
Institutional risk governance
\end{minipage} & \begin{minipage}[t]{\linewidth}\raggedright
Is the institution managing AI risk through sound processes?
\end{minipage} & \begin{minipage}[t]{\linewidth}\raggedright
Enterprise
\end{minipage} & \begin{minipage}[t]{\linewidth}\raggedright
Periodic
\end{minipage} & \begin{minipage}[t]{\linewidth}\raggedright
Board and senior management; auditors and certifiers
\end{minipage} & \begin{minipage}[t]{\linewidth}\raggedright
NIST AI RMF {[}5,6{]}; EU AI Act {[}7{]}; CRI Profile {[}8{]}; ISO/IEC 42001 {[}9{]}; supervisory guidance {[}4{]}
\end{minipage} & \begin{minipage}[t]{\linewidth}\raggedright
Sets risk appetite: which tiers and action classes are acceptable, and what assurance level is required. The Profile is the per-deployment evidence these frameworks call for without specifying.
\end{minipage} \\
\begin{minipage}[t]{\linewidth}\raggedright
\textbf{Deployment disclosure and assurance}
\end{minipage} & \begin{minipage}[t]{\linewidth}\raggedright
\textbf{What is this deployment permitted to do, how autonomously, in what context, under what containment, and how well is that evidenced?}
\end{minipage} & \begin{minipage}[t]{\linewidth}\raggedright
\textbf{Deployment}
\end{minipage} & \begin{minipage}[t]{\linewidth}\raggedright
\textbf{Point in time, with a validity window and invalidation triggers}
\end{minipage} & \begin{minipage}[t]{\linewidth}\raggedright
\textbf{Accountable owner; independent assessor at Level 3}
\end{minipage} & \begin{minipage}[t]{\linewidth}\raggedright
\textbf{The Agent Reliability Profile}
\end{minipage} & \begin{minipage}[t]{\linewidth}\raggedright
\textbf{This paper.}
\end{minipage} \\
\begin{minipage}[t]{\linewidth}\raggedright
Identity and authority
\end{minipage} & \begin{minipage}[t]{\linewidth}\raggedright
Which agent is this, and on whose authority does it act?
\end{minipage} & \begin{minipage}[t]{\linewidth}\raggedright
Agent instance and its credentials
\end{minipage} & \begin{minipage}[t]{\linewidth}\raggedright
Each session or interaction
\end{minipage} & \begin{minipage}[t]{\linewidth}\raggedright
Credential issuer; the relying party verifies
\end{minipage} & \begin{minipage}[t]{\linewidth}\raggedright
Verifiable credentials {[}26{]}; agent identity and authorization standards work {[}23,24{]}
\end{minipage} & \begin{minipage}[t]{\linewidth}\raggedright
Binds the live instance to the profiled deployment, so that a counterparty can confirm the agent in front of it is the one the claim describes.
\end{minipage} \\
\begin{minipage}[t]{\linewidth}\raggedright
Runtime control
\end{minipage} & \begin{minipage}[t]{\linewidth}\raggedright
Should this action proceed, now?
\end{minipage} & \begin{minipage}[t]{\linewidth}\raggedright
Individual action
\end{minipage} & \begin{minipage}[t]{\linewidth}\raggedright
Real time
\end{minipage} & \begin{minipage}[t]{\linewidth}\raggedright
The operator's control plane
\end{minipage} & \begin{minipage}[t]{\linewidth}\raggedright
SAFR {[}16{]}; runtime governance frameworks {[}15{]}; agentic threat models {[}25{]}
\end{minipage} & \begin{minipage}[t]{\linewidth}\raggedright
Enforces the boundary the Profile declares. Its telemetry returns as evidence and as invalidation triggers.
\end{minipage} \\
\midrule\noalign{}
\endhead
\bottomrule\noalign{}
\endlastfoot
\end{longtable}
\endgroup

\hypertarget{model-and-product-evaluation}{%
\subsection{9.1 Model and product evaluation}\label{model-and-product-evaluation}}

Deployed agents are downstream from foundation models and other products that support agentic systems. Evaluations of those upstream models and products can provide some degree of assurance. Model cards, for instance, provide structured reporting of intended use, evaluation conditions, performance, and limitations {[}27{]}, and system cards extend the practice to composed systems. Tool-agent-user interaction benchmarks {[}17{]} and shared evaluation infrastructure {[}18{]} provide task environments and comparability scaffolding, giving users a sense of the abilities and limits of salient models and products.

Evidence at this layer measures performance against tasks, under conditions the developer or the benchmark body chose. The Profile, by contrast, describes and tests a deployment as an institution has configured it. Model and product evidence is, thus, a complementary upstream input. A strong result can reduce what a deployment must demonstrate for itself, and a vendor assessment can sit inside a Profile Package, but neither can substitute for a claim bound to the deployment.

\hypertarget{institutional-risk-governance}{%
\subsection{9.2 Institutional risk governance}\label{institutional-risk-governance}}

AI governance and risk frameworks assess institutions and their risk-management processes, from use-case selection through deployment, monitoring, and reporting. Their unit of analysis is the institution or enterprise. They establish that an institution has the roles, policies, and controls to manage AI risk; they do not assess any particular deployed agent.

\textbf{Table 6}. Institutional governance frameworks and their gaps with respect to agent reliability

\begingroup\footnotesize
\begin{longtable}[]{@{}
  >{\raggedright\arraybackslash}p{(\columnwidth - 4\tabcolsep) * \real{0.2400}}
  >{\raggedright\arraybackslash}p{(\columnwidth - 4\tabcolsep) * \real{0.3800}}
  >{\raggedright\arraybackslash}p{(\columnwidth - 4\tabcolsep) * \real{0.3800}}@{}}
\toprule\noalign{}
\begin{minipage}[t]{\linewidth}\raggedright
Framework
\end{minipage} & \begin{minipage}[t]{\linewidth}\raggedright
Contribution
\end{minipage} & \begin{minipage}[t]{\linewidth}\raggedright
Gap with respect to agent reliability
\end{minipage} \\
\midrule\noalign{}
\begin{minipage}[t]{\linewidth}\raggedright
AI risk-management frameworks (e.g., the NIST AI RMF {[}5,6{]})
\end{minipage} & \begin{minipage}[t]{\linewidth}\raggedright
A governance scaffold: functions to govern, map, measure, and manage, and a vocabulary of trustworthy-AI characteristics.
\end{minipage} & \begin{minipage}[t]{\linewidth}\raggedright
No metric, threshold, or benchmark. Framework compliance leaves agent reliability undetermined. Evidentiary expectations are unspecified. The concepts of agent autonomy, action, and operating domain are absent.
\end{minipage} \\
\begin{minipage}[t]{\linewidth}\raggedright
Risk-tiered AI legislation (e.g., the EU AI Act {[}7{]})
\end{minipage} & \begin{minipage}[t]{\linewidth}\raggedright
Mandatory classification of use cases by risk, with obligations for conformity assessment required for high-risk systems management, data governance, logging, human oversight, and robustness.
\end{minipage} & \begin{minipage}[t]{\linewidth}\raggedright
Classifies use cases and imposes process to operationalize governance; the operational reliability of a deployed agent remains unmeasured. Human oversight is required without a defined autonomy scale.
\end{minipage} \\
\begin{minipage}[t]{\linewidth}\raggedright
Sector control catalogues (e.g., the CRI Profile {[}8{]})
\end{minipage} & \begin{minipage}[t]{\linewidth}\raggedright
Consolidated control objectives mapped to existing regulatory expectations; the reference standard for cyber and operational risk in financial services.
\end{minipage} & \begin{minipage}[t]{\linewidth}\raggedright
The unit of analysis is the enterprise and its controls. Assesses whether controls exist, without observing the live configuration of running systems.
\end{minipage} \\
\begin{minipage}[t]{\linewidth}\raggedright
AI management-system standards (e.g., ISO/IEC 42001 {[}9{]})
\end{minipage} & \begin{minipage}[t]{\linewidth}\raggedright
The closest existing regime to certifiability: an auditable management system for AI, assessed by accredited third parties.
\end{minipage} & \begin{minipage}[t]{\linewidth}\raggedright
Certifies that an institution has a well-structured AI management system. Provides no framework for per-agent unit of evidence and makes no claim about the behavior of any particular agent.
\end{minipage} \\
\midrule\noalign{}
\endhead
\bottomrule\noalign{}
\endlastfoot
\end{longtable}
\endgroup

Governance sets the terms -- e.g., an institution's risk appetite determines how much autonomy and what agentic actions are acceptable in which domains. But governance exists at the level of the institution. The Profile, by contrast, focuses on specific deployments and on evidence in a common vocabulary that the governance process can inventory, challenge, and report. In the U.S. this is particularly salient since the adoption of SR 26-2 by the federal banking agencies, which places generative and agentic AI outside the scope of model risk management guidance and directs institutions to their broader risk-management practices {[}4{]}.

\hypertarget{deployment-disclosure-and-assurance}{%
\subsection{9.3 Deployment disclosure and assurance}\label{deployment-disclosure-and-assurance}}

The focus of this paper on the deployment layer builds on two precedents. The first is structured disclosure. Model and system cards {[}27{]} and FactSheets, which proposed supplier declarations of conformity for AI services {[}38{]}, showed that standardized documentation makes systems comparable and claims contestable. The Profile extends this to deployed agents. The second is graduated assurance, in which the same claim structure can be backed by evidence of increasing rigor and independence, as in the evaluation assurance levels of the Common Criteria {[}28{]}.

Thus, the Profile combines these into a four-axis operating boundary, minimum floors that must be passed regardless of aggregate performance, a production assurance ladder with a separate benchmark comparability flag, and invalidation triggers that tie the Profile's validity to material change. In doing so, it bridges the institutional layers above focused on model and product evaluations and risk governance with the action layers below focused on identity and runtime controls.

\hypertarget{identity-and-authority}{%
\subsection{9.4 Identity and authority}\label{identity-and-authority}}

The identity layer establishes which agent is acting, which principal it acts for, and what authority has been delegated to it, in a form the relying party can verify at the moment of interaction. It is the equivalent of customer due diligence, but for agents, hence the moniker ``know-your-agent''.

A range of standards exists. The W3C Verifiable Credentials data model provides an issuer--holder--verifier structure for tamper-evident, machine-verifiable claims {[}26{]}. The NIST National Cybersecurity Center of Excellence has opened work on identity and authorization for software and AI agents {[}23{]}, and agent identity sits alongside security evaluation and interoperability on the agenda of NIST's AI Agent Standards Initiative {[}24{]}. Enterprise agent platforms and agent security products increasingly issue their own agent identifiers, each under its own schema.

Identity and the Profile are distinct. An identity credential establishes who is acting and with what delegated authority. A Profile, by contrast, establishes the bounded reliability claim.

\hypertarget{runtime-control}{%
\subsection{9.5 Runtime control}\label{runtime-control}}

Runtime controls operate on individual agent actions as they happen. The Monetary Authority of Singapore, with industry partners, published Safeguards for Agentic Finance at Runtime (SAFR) in July 2026: a reference model in which an agent's identity, delegated authority, institutional controls, and risk thresholds are checked at runtime before actions execute, with a disposition engine and a tamper-evident audit log {[}16{]}. Related research frameworks propose trajectory telemetry, continuous authorization, policy-conformance checking, and tier-based containment {[}15{]}{[}25{]}. Some agent builder platforms implement the same functions in their control planes (Section 2.2).

The Profile describes and tests runtime controls; it does not perform them. Conversely, a runtime control plane decides whether an action may proceed; it does not state , in a form a relying party can examine, what the deployment as a whole is permitted to do or how well that permission structure has been evidenced.

The two, therefore, comprise two halves of one mechanism. The Profile's scope, S(D), is a declarative statement of the boundary a runtime control plane should enforce. The control plane, in turn, is the natural source of the operational evidence a Profile cites, and its telemetry can trigger the invalidation and re-profiling events described in Section 5.3. Runtime controls without a Profile enforces limits that no outside party can inspect; a Profile without runtime controls describes limits that cannot be enforced.

\hypertarget{the-layers-in-operation}{%
\subsection{9.7 The layers in operation}\label{the-layers-in-operation}}

Together the five layers described above form a loop. Governance decides what is acceptable. The Profile declares the operating boundary for a deployment and the evidence that the boundary holds. Identity binds that declaration to the running instance. Runtime controls enforce the boundary action by action. What runtime observes flows back, as evidence that the claim continues to hold or as a trigger that voids it, and a voided Profile returns the deployment to governance for a fresh decision.

Consider a payment-exceptions agent at a bank. The vendor's product evaluation shows that the underlying agent handles multi-step reconciliation tasks reliably. The bank's governance process sets its appetite: no autonomous movement of funds, human approval for any customer communication. The deployment's Profile states the resulting claim, at tier T2, within the payment-exceptions domain, for read, analyse, draft, and write-internal action classes, with funds movement and external communication explicitly prohibited, and records that the claim has been validated at Level 2. The agent's credential references that Profile, so that the payments platform it calls can confirm which deployment is calling and what it is cleared to do. The runtime control plane blocks a funds-movement call the agent was never entitled to make and logs the attempt. When the vendor ships a new model version, the change matches an entry on the Profile's invalidation list; the Profile is suspended pending retest, and the governance committee sees that in its inventory the same day.

No single layer produces that outcome. Product evaluation cannot see the bank's configuration; governance cannot see the individual deployment; identity cannot speak to reliability; runtime control cannot explain itself to an outside party. The deployment layer connects them together, operationalized by the Profile.

\hypertarget{limitations-and-open-questions}{%
\section{10 Limitations and open questions}\label{limitations-and-open-questions}}

The proposal carries eight principal limitations.

\emph{Point-in-time evidence.} A Profile describes an agent's configuration at a moment, and agentic systems drift continuously. Invalidation triggers, evidence-freshness fields, monitoring, and other programmatic mitigants can reduce the window of stale assurance, but residual risk likely remains between observations.

\emph{Test-to-production transfer.} Reference environments omit production complexity, a standing weakness of every benchmark that agent evaluations compound {[}19,20{]}. Environment-fidelity statements make the gap visible.

\emph{Self-assurance at the lower levels.} Levels 1 and 2 are self-assurance by design and are labeled as such on the artifact. The value of Level~3 and of the benchmark comparability flag depends on an independence model---assessor qualification, conflict rules, sampling, oversight---that requires institutional design, which is beyond the scope of this paper (Section~\protect\hyperlink{implementation-program-and-research-agenda}{11}).

\emph{Incentive effects.} Institutions and vendors carry incentives toward favorable classification. Countermeasures include: classification by evidence, treatment of unevidenced claims as unprofiled, and behavioral indicators for the autonomy tier. The sufficiency of these can be empirically tested during pilots.

\emph{Human-control measurement.} Approval-log indicators and habituation testing require calibration, labor protections, and safeguards against punitive misuse. The mapping from indicator thresholds to tier classification needs field validation {[}3{]}.

\emph{Open taxonomies.} ODD instances evolve, action-class boundary cases will consume committee time, and the control catalogue requires maintenance. Versioned governance of the taxonomy is necessary throughout.

\emph{Bounded subject.} The Profile measures operational reliability. Fairness, explainability, data governance, consumer compliance, and cybersecurity adequacy remain governed by their own frameworks.

\emph{Platform concentration.} The Platform Baseline Profile concentrates assurance: a defective platform claim undermines every anchored Profile simultaneously, and adversarial testing of platform-derived agents at scale is an open methodological problem (Section \protect\hyperlink{implementation-program-and-research-agenda}{11}).

Different technical environments pose another challenge. Published architectures for deploying agents safely assume that an agent's tools are listed in advance, every use of a tool passes through a control system, and the agent cannot gain new tool access while running {[}31,32{]}. Agents created by a platform have a limited exception: they may select tools while running, but only from a controlled registry. Section \href{https://docs.google.com/document/d/1fObtlznh4eKOsgm-0WDAWhJg4bT_zrYxYAeT4LYwidI/edit\#bookmark=id.no53s3lvuph9}{\ul{2.2}} explains how the framework applies to them. Institutions with a mix of legacy systems may find it hard to list every tool an agent can use. If they cannot produce that list when asked, the deployment should receive close scrutiny, whatever tier it claims {[}3{]}. We invite comment on whether the framework should accommodate this situation or record it as a gap.

\hypertarget{implementation-program-and-research-agenda}{%
\section{11 Implementation program and research agenda}\label{implementation-program-and-research-agenda}}

We propose proceeding through five phases, each with a defined success criterion and, where appropriate, an off-ramp.

\begin{enumerate}
\def\labelenumi{\arabic{enumi}.}
\item
  \begin{quote}
  \emph{Agree on the vocabulary.} Publish the four axes, the autonomy scale, the action ontology, and the control catalogue as a common reference, with a crosswalk to existing frameworks. We expect this to produce robust discussion, not immediate agreement with the contents of the tables in this paper. Success: two institutions describe the same deployment in the same words, tested through inter-rater classification of shared example deployments. Decision point: whether the vocabulary is actionable as written.
  \end{quote}
\item
  \begin{quote}
  \emph{Populate Level 1 Profiles.} Institutions produce Asserted Profiles for real deployments from their own materials, with effort measured. Success: Profiles that businesses and risk committees use, at a credible production cost. Off-ramp: if production is expensive with few economies of scale or the result lacks decision value.
  \end{quote}
\item
  \begin{quote}
  \emph{Test and populate Level 2 Profiles.} The same institutions run the Profile Validator against their own environments. Success: findings that surprise them---controls documented but unconfigured, permissions broader than policy. Decision point: whether the findings justify the testing cost.
  \end{quote}
\item
  \begin{quote}
  \emph{Resolve independence.} Settle what qualifies an independent assessor for Level 3, whether evidence travels or the assessor works on site, retention and conflict rules, and whether a registry samples submitted Profiles.
  \end{quote}
\item
  \begin{quote}
  \emph{Build the shared benchmark environment.} A single reference ODD, co-built with institutions and domain experts, producing the first Benchmarked Profiles and the first comparable results. This is likely the most expensive and technically demanding item {[}1{]}.
  \end{quote}
\end{enumerate}

Going forward, the program should focus on the following research questions:

\begin{enumerate}
\def\labelenumi{\arabic{enumi}.}
\item
  \begin{quote}
  How consistently can different reviewers classify an agent's autonomy level and the types of actions it can take?
  \end{quote}
\item
  \begin{quote}
  Which patterns in approval logs show that a person carefully reviewed a decision, and how should the minimum standard differ by field?
  \end{quote}
\item
  \begin{quote}
  Do tests of reviewer complacency predict how people respond to unusual cases in real operations?
  \end{quote}
\item
  \begin{quote}
  Which significant changes require a full new evaluation, and which need only limited retesting?
  \end{quote}
\item
  \begin{quote}
  How many times should a test be repeated when an agent's behavior varies or errors could have serious consequences?
  \end{quote}
\item
  \begin{quote}
  Which safeguards keep benchmarks trustworthy when agents can recognize that they are being tested?
  \end{quote}
\item
  \begin{quote}
  How can evidence from reusable components be combined without overstating confidence in the complete deployment?
  \end{quote}
\item
  \begin{quote}
  Which Profile details can remain private while still allowing others to verify the findings?
  \end{quote}
\item
  \begin{quote}
  How should assurance levels reflect deployment risk and guide supervisory attention?
  \end{quote}
\item
  \begin{quote}
  How can challenging tests across a large fleet of agents check a platform's claim that its safeguards apply to every agent?
  \end{quote}
\item
  \begin{quote}
  Can records of each session's configuration reliably show, afterward, which permissions and safeguards were in effect?
  \end{quote}
\item
  \begin{quote}
  Which Profile measures best predict real incidents and how well agents recover from them?
  \end{quote}
\end{enumerate}

We strongly encourage feedback from stakeholders. From institutions, what was most challenging when producing a full Profile for a real agentic deployment? From supervisors, is the bounded claim the right thing to prioritize and are the four axes the right things to ask about? From assurance firms, what are your perspectives on the Level~3 qualified-independent-assessor role from practical, commercial, and governance perspectives? In general, will the pace and direction of agent development render this approach obsolete, and if so what should replace it? We are most interested in divergent perspectives.

\hypertarget{conclusion}{%
\section{12 Conclusion}\label{conclusion}}

Agentic systems create a deployment-level information problem: reliability depends on autonomy, operating context, action authority, and layered containment. The Agent Reliability Profile builds a shared vocabulary for populating agent artifacts that facilitates inter- and intra-organizational understanding of reliability. The four axes of the Agent Reliability Profile define a bounded, falsifiable claim; its floors provide a minimum standard; its provenance map and Profile Package support examination; its invalidation list ties assurance to change; and its three production assurance levels let institutions begin with assertion-based inventory and progress through in-environment validation and independent verification, while a separate benchmark comparability flag supports comparison across institutions, with the Profile Builder, Profile Validator, and Profile Benchmarker operationalizing the ladder. The next step is empirical: compiling Profiles for real deployments, comparing independent classifications, running the Validator against live configurations, and observing impacts on decisions. To become a standard the Profile needs to be used repeatedly, have predictive utility, and be revised in response to failures. The architecture proposed here offers a concrete basis for that work.

\hypertarget{acknowledgments}{%
\section{Acknowledgments}\label{acknowledgments}}

The authors thank Sean McGregor (MLCommons) and Deborah Eng (MLCommons; JPMorgan Chase) for their contributions to this work, and participants in the MLCommons Financial Services Working Group and the broader AI risk, reliability, evaluation, and standards communities whose discussions informed this proposal.

During the preparation of this work, the authors used Claude Opus 5 and GPT 5.6-Sol to assist with some drafting. After using the tools, the authors reviewed and edited the content and take full responsibility for all aspects of the publication.

\hypertarget{appendix-a.-the-bounded-claim}{%
\section{Appendix A. The bounded claim}\label{appendix-a.-the-bounded-claim}}

Below are two illustrative examples.

\hypertarget{a.1-a-coding-agent}{%
\subsection{A.1 A coding agent}\label{a.1-a-coding-agent}}

\begin{quote}
\emph{Reliable at autonomy tier T2, within the internal software-development operational design domain, for action classes \{A1 read/retrieve, A2 analyse/compute, A3 draft, selected A5 write internal: open a pull request\}, under \{human approval to merge, least-privilege repository token, full execution logging\}.}
\end{quote}

\emph{Autonomy tier.} The agent plans and executes multi-step coding work and opens pull requests. A person approves any merge to a protected branch and any deployment.

\emph{Operational design domain.} Internal tier-2 services under the institution's change-management regime: the repositories, build pipeline, issue tracker, review process, branch-protection rules, and test suites in scope.

\emph{Permitted actions.} Read code, run tests, draft code, open and comment on a pull request.

\emph{Prohibited actions.} Merge, deploy, alter branch protection or access controls, change pipeline configuration, access secrets.

\emph{Floors.} No attempted catastrophic action during the run. Boundary detection above the mandatory threshold. No unrecoverable tool-state failures.

\emph{Out of scope.} Behavior outside this domain or above this tier remains unattested. Use on customer-facing systems, on changes touching access control or production deployment, or under any looser control envelope requires separate assessment.

\emph{Validity.} One month, or any material change to the model, the tooling, or the control envelope, whichever comes first.

\hypertarget{a.2-a-financial-crime-alert-triage-agent}{%
\subsection{A.2 A financial-crime alert-triage agent}\label{a.2-a-financial-crime-alert-triage-agent}}

\begin{quote}
\emph{Reliable at autonomy tier T2, within the retail alert-triage operational design domain, for action classes \{A1 read/retrieve, A2 analyse/compute, A3 draft narrative, A4 recommend\}, under \{analyst approval before any disposition, read-only access to customer data, evidence-preserving execution record\}.}
\end{quote}

\emph{Autonomy tier.} The agent works an alert end to end and produces a disposition recommendation with its reasoning. An analyst approves before any case is dispositioned. Closing a case and making a regulatory filing remain outside the agent's authority.

\emph{Operational design domain.} Retail transaction-monitoring alerts under the institution's financial-crime programme, scoped to structuring and rapid-movement-of-funds typologies: the case-management system, transaction histories, customer profiles, screening data, typology playbooks, and the institution's filing standard.

\emph{Permitted actions.} Read case data, query transaction history, retrieve policy and typology guidance, draft a narrative, recommend a disposition.

\emph{Prohibited actions.} Make a regulatory filing, close or disposition a case, notify a customer, freeze an account, contact a third party, share customer data outside the case context.

\emph{Floors.} No attempted prohibited disposition, filing, or customer contact. No fabricated facts in a drafted narrative---the hard safety floor for this domain. Boundary detection above the mandatory threshold. Verified handling of customer data.

\emph{Out of scope.} Typologies outside the assessed set---trade-based laundering, terrorist financing, sanctions evasion, complex shell structures---fall outside this claim, as do other jurisdictions and any configuration that loosens the stated envelope.

\emph{Validity.} One month, or any material change to the model, the tooling, the typology scope, or the control envelope.

The two examples share a structure. The prohibited list carries as much of the claim as the permitted list; the floors are conditions rather than scores; the out-of-scope statement is longer than the claim; and the validity window is keyed to change rather than to the calendar.

\hypertarget{appendix-b.-the-full-option-universe-of-the-four-axes}{%
\section{Appendix B. The full option universe of the four axes}\label{appendix-b.-the-full-option-universe-of-the-four-axes}}

This appendix summarizes the full option universe of the four axes as currently mapped by the working group, and reproduces the companion one-pager as Exhibits B1--B3 {[}34{]}: five autonomy tiers and the residual status U, with six cross-cutting modifiers; an open set of operational design domains illustrated by fifty-two candidate use cases and additional multiplying factors; twelve action classes; and fifty-five named controls across six layers.

\hypertarget{b.1-autonomy-tier-five-tiers-and-a-residual-status-six-modifiers}{%
\subsection{B.1 Autonomy tier: five tiers and a residual status, six modifiers}\label{b.1-autonomy-tier-five-tiers-and-a-residual-status-six-modifiers}}

The five tiers and the residual status U appear in Table~\protect\hyperlink{tab1}{1}. Tiers attach to deployments and are defined by evidence. Two structural watchpoints organize testing attention: the T2/T3 boundary (substantive review as against reflexive approval---a behavioral distinction) and the T3/T4 boundary (human fallback as against tested safe state). The six cross-cutting modifiers---data authority, reversibility profile, customer and market impact, stack composability, integrated-stack accountability, and population scope---record risk dimensions the tier alone leaves unexpressed {[}3{]}.

\hypertarget{b.2-operational-design-domain-an-open-set}{%
\subsection{B.2 Operational design domain: an open set}\label{b.2-operational-design-domain-an-open-set}}

Fifty-two candidate use cases in financial services, listed to show the shape of the axis.

\emph{Retail and consumer banking (8).} Servicing chat; complaint handling; dispute resolution; account opening and onboarding; fraud alert triage; collections and hardship; personal financial guidance; deposit operations exceptions.

\emph{Lending and credit (7).} Consumer underwriting; mortgage origination; commercial underwriting and spreading; credit-line management; covenant monitoring; loan servicing and modifications; appraisal and collateral review.

\emph{Payments (4).} Payment-operations exceptions and wire repair; screening alert triage; chargeback and representment processing; real-time fraud interdiction.

\emph{Financial crime and compliance (10).} Alert triage for structuring and rapid movement; alert triage for trade-based laundering, terrorist financing, sanctions evasion, and shell structures; periodic customer review and enhanced diligence; narrative drafting and quality assurance; monitoring-rule tuning; trade surveillance; communications surveillance; regulatory-change monitoring; marketing and conduct review; conflicts monitoring.

\emph{Markets (6).} Research generation; execution and algorithm selection; post-trade reconciliation and settlement exceptions; corporate actions; margin calls; institutional onboarding.

\emph{Wealth and asset management (5).} Portfolio rebalancing; proposal generation; suitability review; client reporting; tax-loss harvesting.

\emph{Insurance and treasury (5).} Claims triage and adjudication; underwriting; liquidity forecasting; financial close and reconciliation; regulatory reporting preparation.

\emph{Internal functions (7).} Software development; IT operations and incident response; security operations triage; model validation support; internal audit support; procurement and third-party review; employee services.

Each use case multiplies by the customer population it serves, the sensitivity of the data it touches, the cost of failure, and the jurisdiction. The ODD axis admits standardization of its schema only; instances are built one at a time by the institutions developing agents.

\hypertarget{b.3-action-class-twelve-classes}{%
\subsection{B.3 Action class: twelve classes}\label{b.3-action-class-twelve-classes}}

The twelve classes are graded by reversibility and consequence from A1 (read, no side effects) to A12 (delegate and spawn, amplifying). A Profile fixes a permitted set plus an explicit prohibited set; the axis value is a pair of sets. Emergent composite actions in multi-agent systems can add capabilities that no single agent holds, and are tested at the system level.

Table 2. The twelve action classes, graded by reversibility and consequence.

\begingroup\footnotesize
\begin{longtable}[]{@{}
  >{\raggedright\arraybackslash}p{(\columnwidth - 6\tabcolsep) * \real{0.0799}}
  >{\raggedright\arraybackslash}p{(\columnwidth - 6\tabcolsep) * \real{0.2400}}
  >{\raggedright\arraybackslash}p{(\columnwidth - 6\tabcolsep) * \real{0.4300}}
  >{\raggedright\arraybackslash}p{(\columnwidth - 6\tabcolsep) * \real{0.2501}}@{}}
\toprule\noalign{}
\begin{minipage}[t]{\linewidth}\raggedright
\end{minipage} & \begin{minipage}[t]{\linewidth}\raggedright
Class
\end{minipage} & \begin{minipage}[t]{\linewidth}\raggedright
Examples
\end{minipage} & \begin{minipage}[t]{\linewidth}\raggedright
How hard to undo
\end{minipage} \\
\midrule\noalign{}
\begin{minipage}[t]{\linewidth}\raggedright
A1
\end{minipage} & \begin{minipage}[t]{\linewidth}\raggedright
Read / retrieve
\end{minipage} & \begin{minipage}[t]{\linewidth}\raggedright
Query transactions, read case files, search policy
\end{minipage} & \begin{minipage}[t]{\linewidth}\raggedright
n/a\footnote{This ignores the risk of data leakage or theft due from cyber vulnerabilities, which are out of scope for the Profile.}
\end{minipage} \\
\begin{minipage}[t]{\linewidth}\raggedright
A2
\end{minipage} & \begin{minipage}[t]{\linewidth}\raggedright
Analyse / compute
\end{minipage} & \begin{minipage}[t]{\linewidth}\raggedright
Score, simulate, reconcile in memory
\end{minipage} & \begin{minipage}[t]{\linewidth}\raggedright
n/a
\end{minipage} \\
\begin{minipage}[t]{\linewidth}\raggedright
A3
\end{minipage} & \begin{minipage}[t]{\linewidth}\raggedright
Draft for a person
\end{minipage} & \begin{minipage}[t]{\linewidth}\raggedright
Narratives, code, emails, memos, draft filings
\end{minipage} & \begin{minipage}[t]{\linewidth}\raggedright
Fully reversible
\end{minipage} \\
\begin{minipage}[t]{\linewidth}\raggedright
A4
\end{minipage} & \begin{minipage}[t]{\linewidth}\raggedright
Recommend / route
\end{minipage} & \begin{minipage}[t]{\linewidth}\raggedright
Disposition recommendations, escalations, prioritisation
\end{minipage} & \begin{minipage}[t]{\linewidth}\raggedright
Fully reversible
\end{minipage} \\
\begin{minipage}[t]{\linewidth}\raggedright
A5
\end{minipage} & \begin{minipage}[t]{\linewidth}\raggedright
Write internal
\end{minipage} & \begin{minipage}[t]{\linewidth}\raggedright
Case notes, tickets, opening a pull request, CRM updates
\end{minipage} & \begin{minipage}[t]{\linewidth}\raggedright
Reversible
\end{minipage} \\
\begin{minipage}[t]{\linewidth}\raggedright
A6
\end{minipage} & \begin{minipage}[t]{\linewidth}\raggedright
Submit / commit internal
\end{minipage} & \begin{minipage}[t]{\linewidth}\raggedright
Merge code, close a case,
\end{minipage} & \begin{minipage}[t]{\linewidth}\raggedright
Semi-reversible
\end{minipage} \\
\begin{minipage}[t]{\linewidth}\raggedright
A7
\end{minipage} & \begin{minipage}[t]{\linewidth}\raggedright
External communication
\end{minipage} & \begin{minipage}[t]{\linewidth}\raggedright
Notify a customer, contact a third party, publish
\end{minipage} & \begin{minipage}[t]{\linewidth}\raggedright
Hard to reverse
\end{minipage} \\
\begin{minipage}[t]{\linewidth}\raggedright
A8
\end{minipage} & \begin{minipage}[t]{\linewidth}\raggedright
Transact---bounded
\end{minipage} & \begin{minipage}[t]{\linewidth}\raggedright
Place a hold, block a card, trade within a cancel window
\end{minipage} & \begin{minipage}[t]{\linewidth}\raggedright
Partially reversible
\end{minipage} \\
\begin{minipage}[t]{\linewidth}\raggedright
A9
\end{minipage} & \begin{minipage}[t]{\linewidth}\raggedright
Move funds
\end{minipage} & \begin{minipage}[t]{\linewidth}\raggedright
Wire release, payment execution, settlement
\end{minipage} & \begin{minipage}[t]{\linewidth}\raggedright
Irreversible
\end{minipage} \\
\begin{minipage}[t]{\linewidth}\raggedright
A10
\end{minipage} & \begin{minipage}[t]{\linewidth}\raggedright
Regulatory filing
\end{minipage} & \begin{minipage}[t]{\linewidth}\raggedright
File a suspicious-activity or currency report, submit returns, attest
\end{minipage} & \begin{minipage}[t]{\linewidth}\raggedright
Irreversible, with legal effect
\end{minipage} \\
\begin{minipage}[t]{\linewidth}\raggedright
A11
\end{minipage} & \begin{minipage}[t]{\linewidth}\raggedright
Access and infrastructure change
\end{minipage} & \begin{minipage}[t]{\linewidth}\raggedright
Alter permissions, deploy to production, change branch protection
\end{minipage} & \begin{minipage}[t]{\linewidth}\raggedright
Irreversible and risk-amplifying
\end{minipage} \\
\begin{minipage}[t]{\linewidth}\raggedright
A12
\end{minipage} & \begin{minipage}[t]{\linewidth}\raggedright
Delegate / spawn
\end{minipage} & \begin{minipage}[t]{\linewidth}\raggedright
Create sub-agents, grant credentials to another agent
\end{minipage} & \begin{minipage}[t]{\linewidth}\raggedright
Amplifying
\end{minipage} \\
\midrule\noalign{}
\endhead
\bottomrule\noalign{}
\endlastfoot
\end{longtable}
\endgroup

\hypertarget{b.4-control-envelope-fifty-five-controls-across-six-layers}{%
\subsection{B.4 Control envelope: fifty-five controls across six layers}\label{b.4-control-envelope-fifty-five-controls-across-six-layers}}

An envelope is a subset of the catalogue below; in practice, controls bundle into a small number of standard per-layer profiles. The catalogue mirrors existing control frameworks and can be assembled by crosswalk.

\begingroup\footnotesize
\begin{longtable}[]{@{}
  >{\raggedright\arraybackslash}p{(\columnwidth - 2\tabcolsep) * \real{0.2000}}
  >{\raggedright\arraybackslash}p{(\columnwidth - 2\tabcolsep) * \real{0.8000}}@{}}
\toprule\noalign{}
\begin{minipage}[t]{\linewidth}\raggedright
\textbf{Layer}
\end{minipage} & \begin{minipage}[t]{\linewidth}\raggedright
\textbf{Controls}
\end{minipage} \\
\midrule\noalign{}
\begin{minipage}[t]{\linewidth}\raggedright
Model (9)
\end{minipage} & \begin{minipage}[t]{\linewidth}\raggedright
Structured plan review before execution; uncertainty surfacing and confidence thresholds; irreversible-action flagging; evidence grounding (cite or refuse); out-of-scope flagging; refusal policies for prohibited requests; self-verification pass before output; system-prompt hardening; constraints on customer-facing tone and claims.
\end{minipage} \\
\begin{minipage}[t]{\linewidth}\raggedright
Tool (10)
\end{minipage} & \begin{minipage}[t]{\linewidth}\raggedright
Least-privilege credentials; permitted and prohibited action allowlists; read-only scoping of data access; per-case or per-session data scoping; transaction caps by value, volume, and velocity; time-boxed credentials; dual control enforced at the tool; redaction or tokenisation at the boundary; no-tool-available design, where the capability is simply absent; structured schemas constraining free-text arguments.
\end{minipage} \\
\begin{minipage}[t]{\linewidth}\raggedright
Harness (11)
\end{minipage} & \begin{minipage}[t]{\linewidth}\raggedright
Human approval gates by action, checkpoint, or batch; immutable logging of every tool call; evidence-preserving provenance capture; mid-task interruption hooks; automatic routing of high-risk cases to senior review; session budgets in steps, time, and spend; real-time anomaly detection on agent behaviour; sampling-based quality review of outputs; a tested kill switch; rate limiting across sessions; independent second-model verification.
\end{minipage} \\
\begin{minipage}[t]{\linewidth}\raggedright
Environment (9)
\end{minipage} & \begin{minipage}[t]{\linewidth}\raggedright
Sandboxing and segregation from production; network egress policy; hard execution boundaries; rollback or snapshot on any committed change; shadow-mode or canary deployment; encryption at rest and in transit with retention controls; access logging and data-loss prevention; immutable run records; separation of test and live credentials.
\end{minipage} \\
\begin{minipage}[t]{\linewidth}\raggedright
Inter-agent (8)
\end{minipage} & \begin{minipage}[t]{\linewidth}\raggedright
Trust boundaries between agents; message validation and signing; a delegation permission schema; credential isolation and non-propagation; cascade limits; blast-radius caps; topology declaration; sub-agent spawning rules.
\end{minipage} \\
\begin{minipage}[t]{\linewidth}\raggedright
Control plane (8)
\end{minipage} & \begin{minipage}[t]{\linewidth}\raggedright
Registration and ownership of agents, skills, and tools; versioned change records; required evaluation before publication; human review of agent-proposed changes; staged rollout and rollback; deprecation and retirement; change propagation and downstream Profile invalidation; catalogue admission standards for tools.
\end{minipage} \\
\midrule\noalign{}
\endhead
\bottomrule\noalign{}
\endlastfoot
\end{longtable}
\endgroup

\hypertarget{appendix-c.-crosswalk-to-existing-frameworks}{%
\section{Appendix C. Crosswalk to existing frameworks}\label{appendix-c.-crosswalk-to-existing-frameworks}}

Profiles are designed as complements to existing frameworks: an agent assessment should be capable of discharging evidence obligations under several regimes at once.

Table 7. Crosswalk from Profile elements to four families of existing frameworks.

\begingroup\footnotesize
\begin{longtable}[]{@{}
  >{\raggedright\arraybackslash}p{(\columnwidth - 8\tabcolsep) * \real{0.1600}}
  >{\raggedright\arraybackslash}p{(\columnwidth - 8\tabcolsep) * \real{0.2100}}
  >{\raggedright\arraybackslash}p{(\columnwidth - 8\tabcolsep) * \real{0.2100}}
  >{\raggedright\arraybackslash}p{(\columnwidth - 8\tabcolsep) * \real{0.2100}}
  >{\raggedright\arraybackslash}p{(\columnwidth - 8\tabcolsep) * \real{0.2100}}@{}}
\toprule\noalign{}
\begin{minipage}[t]{\linewidth}\raggedright
\textbf{Profile element}
\end{minipage} & \begin{minipage}[t]{\linewidth}\raggedright
\textbf{AI risk-management frameworks}
\end{minipage} & \begin{minipage}[t]{\linewidth}\raggedright
\textbf{Risk-tiered AI legislation}
\end{minipage} & \begin{minipage}[t]{\linewidth}\raggedright
\textbf{Sector control catalogues}
\end{minipage} & \begin{minipage}[t]{\linewidth}\raggedright
\textbf{AI management-system standards}
\end{minipage} \\
\midrule\noalign{}
\begin{minipage}[t]{\linewidth}\raggedright
Autonomy tier
\end{minipage} & \begin{minipage}[t]{\linewidth}\raggedright
Govern---oversight roles
\end{minipage} & \begin{minipage}[t]{\linewidth}\raggedright
Human-oversight obligations
\end{minipage} & \begin{minipage}[t]{\linewidth}\raggedright
Govern function
\end{minipage} & \begin{minipage}[t]{\linewidth}\raggedright
Management-system roles and oversight
\end{minipage} \\
\begin{minipage}[t]{\linewidth}\raggedright
Operational design domain
\end{minipage} & \begin{minipage}[t]{\linewidth}\raggedright
Map---context; Measure
\end{minipage} & \begin{minipage}[t]{\linewidth}\raggedright
Intended purpose; robustness obligations
\end{minipage} & \begin{minipage}[t]{\linewidth}\raggedright
Identify---asset and context
\end{minipage} & \begin{minipage}[t]{\linewidth}\raggedright
AI system impact assessment
\end{minipage} \\
\begin{minipage}[t]{\linewidth}\raggedright
Action class
\end{minipage} & \begin{minipage}[t]{\linewidth}\raggedright
Manage---risk treatment
\end{minipage} & \begin{minipage}[t]{\linewidth}\raggedright
Risk classification; high-risk annexes
\end{minipage} & \begin{minipage}[t]{\linewidth}\raggedright
Protect---entitlements
\end{minipage} & \begin{minipage}[t]{\linewidth}\raggedright
Annex controls
\end{minipage} \\
\begin{minipage}[t]{\linewidth}\raggedright
Control envelope
\end{minipage} & \begin{minipage}[t]{\linewidth}\raggedright
Govern and Manage
\end{minipage} & \begin{minipage}[t]{\linewidth}\raggedright
Risk management, logging, human oversight
\end{minipage} & \begin{minipage}[t]{\linewidth}\raggedright
Protect / Detect / third-party function
\end{minipage} & \begin{minipage}[t]{\linewidth}\raggedright
Operational controls
\end{minipage} \\
\begin{minipage}[t]{\linewidth}\raggedright
The Profile itself
\end{minipage} & \begin{minipage}[t]{\linewidth}\raggedright
Measure---the evidence the framework leaves unspecified
\end{minipage} & \begin{minipage}[t]{\linewidth}\raggedright
Logging and technical-documentation evidence
\end{minipage} & \begin{minipage}[t]{\linewidth}\raggedright
Detect / Respond
\end{minipage} & \begin{minipage}[t]{\linewidth}\raggedright
Documented information
\end{minipage} \\
\midrule\noalign{}
\endhead
\bottomrule\noalign{}
\endlastfoot
\end{longtable}
\endgroup

The specific instruments in view are the NIST AI Risk Management Framework and its generative-AI profile {[}5,6{]}, the EU AI Act and its high-risk obligations {[}7{]}, the Cyber Risk Institute Profile {[}8{]}, and ISO/IEC 42001 {[}9{]}. Jurisdiction-specific mappings---including to SR 26-2, which placed generative and agentic AI outside the scope of United States model-risk guidance and directed institutions to their broader risk-management practices {[}4{]}---form part of the crosswalk deliverable and are deferred from the body of this paper.

\hypertarget{appendix-d.-glossary}{%
\section{Appendix D. Glossary}\label{appendix-d.-glossary}}

\begingroup\footnotesize
\begin{longtable}[]{@{}
  >{\raggedright\arraybackslash}p{(\columnwidth - 2\tabcolsep) * \real{0.2600}}
  >{\raggedright\arraybackslash}p{(\columnwidth - 2\tabcolsep) * \real{0.7400}}@{}}
\toprule\noalign{}
\begin{minipage}[t]{\linewidth}\raggedright
\textbf{Term}
\end{minipage} & \begin{minipage}[t]{\linewidth}\raggedright
\textbf{Meaning}
\end{minipage} \\
\midrule\noalign{}
\begin{minipage}[t]{\linewidth}\raggedright
Agent Reliability Profile
\end{minipage} & \begin{minipage}[t]{\linewidth}\raggedright
The single artifact recording the four axis values, the floors, the out-of-scope and validity statements, the provenance map, and pointers to the underlying evidence. Written in full; each agent has its own Profile.
\end{minipage} \\
\begin{minipage}[t]{\linewidth}\raggedright
Bounded claim
\end{minipage} & \begin{minipage}[t]{\linewidth}\raggedright
The scoped, falsifiable assertion produced by fixing the four axes: reliable at this autonomy tier, in this operational design domain, for these action classes, under this control envelope---and nothing else.
\end{minipage} \\
\begin{minipage}[t]{\linewidth}\raggedright
Autonomy tier
\end{minipage} & \begin{minipage}[t]{\linewidth}\raggedright
How much the agent decides and acts without a person, and who provides the fallback when it cannot continue. The tier determines what counts as a failure and how failures are weighted in evaluation.
\end{minipage} \\
\begin{minipage}[t]{\linewidth}\raggedright
Operational design domain (ODD)
\end{minipage} & \begin{minipage}[t]{\linewidth}\raggedright
The defined operating context---use case, data, tool surface, customer population, cost of failure---that specifies the test environment. Follows its use in automated-driving standards.
\end{minipage} \\
\begin{minipage}[t]{\linewidth}\raggedright
Boundary detection
\end{minipage} & \begin{minipage}[t]{\linewidth}\raggedright
Whether the agent recognises that it has left its tested domain and stops. A defined test at Level 2 and above, distinct from following an escalation rule the agent was given.
\end{minipage} \\
\begin{minipage}[t]{\linewidth}\raggedright
Action class
\end{minipage} & \begin{minipage}[t]{\linewidth}\raggedright
The type and reversibility of actions the agent may take, from read and draft through submit, transact, and move funds. Recorded as a permitted set and a prohibited set.
\end{minipage} \\
\begin{minipage}[t]{\linewidth}\raggedright
Control envelope
\end{minipage} & \begin{minipage}[t]{\linewidth}\raggedright
The containment around the agent---approval gates, least-privilege access, rollback, logging, tested safe state---that makes a given autonomy tier acceptable within a given domain.
\end{minipage} \\
\begin{minipage}[t]{\linewidth}\raggedright
Floor
\end{minipage} & \begin{minipage}[t]{\linewidth}\raggedright
A pass-or-fail condition that cannot be offset by good results elsewhere. Floors are evaluated before any graded result is computed.
\end{minipage} \\
\begin{minipage}[t]{\linewidth}\raggedright
Provenance map
\end{minipage} & \begin{minipage}[t]{\linewidth}\raggedright
A field-by-field record of whether each entry on a Profile was measured, checked against configuration, attested by a person, or is absent.
\end{minipage} \\
\begin{minipage}[t]{\linewidth}\raggedright
Invalidation list
\end{minipage} & \begin{minipage}[t]{\linewidth}\raggedright
The machine-readable statement of which changes void a Profile, which require partial retesting, and which are cosmetic.
\end{minipage} \\
\begin{minipage}[t]{\linewidth}\raggedright
Platform Baseline Profile
\end{minipage} & \begin{minipage}[t]{\linewidth}\raggedright
The Profile carried by an agent platform itself. Its claim is an enforcement claim---the ceilings and invariants the substrate guarantees for every session it hosts---not an attestation of any hosted agent's performance.
\end{minipage} \\
\begin{minipage}[t]{\linewidth}\raggedright
Platform anchor
\end{minipage} & \begin{minipage}[t]{\linewidth}\raggedright
The machine-readable record, on an agent's Profile, of the parent Platform Baseline Profile and version whose substrate controls form part of the agent's control envelope. An anchored Profile is an ordinary Agent Reliability Profile and is void if the parent claim is void.
\end{minipage} \\
\begin{minipage}[t]{\linewidth}\raggedright
Autonomy ceiling
\end{minipage} & \begin{minipage}[t]{\linewidth}\raggedright
The maximum autonomy tier a platform can grant to any agent it hosts, enforced in software and recorded on the platform's Profile in place of a tier.
\end{minipage} \\
\begin{minipage}[t]{\linewidth}\raggedright
Resolved manifest
\end{minipage} & \begin{minipage}[t]{\linewidth}\raggedright
The per-session record of the skills, tools, data contexts, and permissions actually granted to a platform-instantiated agent, resolved from a closed registry and preserved as evidence.
\end{minipage} \\
\begin{minipage}[t]{\linewidth}\raggedright
Profile Package
\end{minipage} & \begin{minipage}[t]{\linewidth}\raggedright
The Profile plus the signed inputs, evidence, and trace it was compiled from. The Profile is the light, scannable summary; the Package is its full backing.
\end{minipage} \\
\begin{minipage}[t]{\linewidth}\raggedright
Asserted Profile (Level 1)
\end{minipage} & \begin{minipage}[t]{\linewidth}\raggedright
Self-reported: compiled by the Profile Builder from the institution's own documents and settings; nothing executed; signed by a named owner.
\end{minipage} \\
\begin{minipage}[t]{\linewidth}\raggedright
Validated Profile (Level 2)
\end{minipage} & \begin{minipage}[t]{\linewidth}\raggedright
Self-reported and self-tested: produced by the Profile Validator run in-house, with reproducible, hashed, signed inputs.
\end{minipage} \\
\begin{minipage}[t]{\linewidth}\raggedright
Verified Profile (Level 3)
\end{minipage} & \begin{minipage}[t]{\linewidth}\raggedright
Self-reported and independently tested: the Profile Validator operated and signed by a qualified independent assessor.
\end{minipage} \\
\begin{minipage}[t]{\linewidth}\raggedright
Benchmarked Profile (benchmark comparability flag)
\end{minipage} & \begin{minipage}[t]{\linewidth}\raggedright
A Profile carrying the benchmark comparability flag: run by the Profile Benchmarker in a shared environment with a hidden, rotating scenario set; comparable across institutions under the reference conditions. A separate field, not an assurance level; it does not raise the Profile's production assurance level.
\end{minipage} \\
\midrule\noalign{}
\endhead
\bottomrule\noalign{}
\endlastfoot
\end{longtable}
\endgroup

\hypertarget{references}{%
\section{References}\label{references}}

{[}1{]} S. McGregor, D. Nathani, L. Saouma, F. Barez, et al. Agentic product maturity ladder V0.1. MLCommons AI Risk and Reliability Working Group, December 2025. \url{https://mlcommons.org/wp-content/uploads/2025/12/ReleaseVersion.pdf}

{[}2{]} S. Rabanser, S. Kapoor, P. Kirgis, K. Liu, S. Utpala, and A. Narayanan. Towards a science of AI agent reliability. \emph{arXiv preprint arXiv:2602.16666}, 2026. \url{https://arxiv.org/abs/2602.16666}

{[}3{]} MLCommons Financial Services Working Group. An autonomy taxonomy for agentic AI in financial services. Discussion paper, version 2, May 2026. Available from the working group on request.

{[}4{]} Board of Governors of the Federal Reserve System, Federal Deposit Insurance Corporation, and Office of the Comptroller of the Currency. SR 26-2: Supervisory guidance on model risk management. April 17, 2026. \url{https://www.federalreserve.gov/supervisionreg/srletters/}

{[}5{]} National Institute of Standards and Technology. Artificial Intelligence Risk Management Framework (AI RMF 1.0), NIST AI 100-1, January 2023. \url{https://doi.org/10.6028/NIST.AI.100-1}

{[}6{]} National Institute of Standards and Technology. Artificial Intelligence Risk Management Framework: Generative Artificial Intelligence Profile, NIST AI 600-1, July 2024. \url{https://doi.org/10.6028/NIST.AI.600-1}

{[}7{]} European Parliament and Council of the European Union. Regulation (EU) 2024/1689 laying down harmonised rules on artificial intelligence (AI Act). \emph{Official Journal of the European Union}, July 2024. \url{https://eur-lex.europa.eu/eli/reg/2024/1689/oj}

{[}8{]} Cyber Risk Institute. The CRI Profile. \url{https://cyberriskinstitute.org/the-profile/}

{[}9{]} International Organization for Standardization and International Electrotechnical Commission. ISO/IEC 42001:2023, Information technology---Artificial intelligence---Management system, December 2023. \url{https://www.iso.org/standard/44545.html}

{[}10{]} SAE International. J3016: Taxonomy and definitions for terms related to driving automation systems for on-road motor vehicles, J3016\_202104, April 2021. \url{https://www.sae.org/standards/content/j3016_202104/}

{[}11{]} European Union Aviation Safety Agency. EASA Artificial Intelligence Concept Paper, Issues 1--3, 2021--2025. \url{https://www.easa.europa.eu/en/domains/research-innovation/ai}

{[}12{]} International Organization for Standardization. ISO 10218: Robotics---Safety requirements for robot systems in an industrial environment. \url{https://www.iso.org/standard/73933.html}

{[}13{]} ASTM International. F3269: Standard practice for methods to safely bound behavior of aircraft systems containing complex functions using run-time assurance. \url{https://www.astm.org/f3269-21.html}

{[}14{]} U.S. Department of Defense. Directive 3000.09: Autonomy in weapons systems. \url{https://www.esd.whs.mil/portals/54/documents/dd/issuances/dodd/300009p.pdf}

{[}15{]} L. Szpruch, A. Sudjianto, T. Bhatti, and G. Ang. Scalable runtime governance for agentic AI in financial services. SSRN 6567199, April 2026. \url{https://doi.org/10.2139/ssrn.6567199}

{[}16{]} Monetary Authority of Singapore and industry partners. Safeguards for Agentic Finance at Runtime (SAFR). July 2026. \url{https://www.mas.gov.sg/publications/monographs-or-information-paper/2026/safeguards-for-agentic-finance-at-runtime}

{[}17{]} S. Yao, N. Shinn, P. Razavi, and K. Narasimhan. \(\tau\)-bench: A benchmark for tool-agent-user interaction in real-world domains. \emph{arXiv preprint arXiv:2406.12045}, 2024. \url{https://arxiv.org/abs/2406.12045}

{[}18{]} S. Kapoor, B. Stroebl, P. Kirgis, et al. Holistic agent leaderboard: The missing infrastructure for AI agent evaluation. Princeton Language and Intelligence, 2025. \url{https://github.com/princeton-pli/hal-harness}

{[}19{]} Y. Zhu et al. Establishing best practices for building rigorous agentic benchmarks. \emph{arXiv preprint arXiv:2507.02825}, 2025. \url{https://arxiv.org/abs/2507.02825}

{[}20{]} M. Hamin and B. Edelman. Cheating on AI agent evaluations. Center for AI Standards and Innovation, National Institute of Standards and Technology, December 2025. \url{https://www.nist.gov/caisi/cheating-ai-agent-evaluations}

{[}21{]} National Institute of Standards and Technology. Challenges to the monitoring of deployed AI systems, NIST AI 800-4, March 2026. \url{https://nvlpubs.nist.gov/nistpubs/ai/NIST.AI.800-4.pdf}

{[}22{]} J. Riggs, M. Hamin, N. Perry, B. Edelman, and P. Cihon. Summary analysis of responses to the request for information regarding security considerations for AI agents, NIST AI 800-5, May 2026. \url{https://www.nist.gov/publications}

{[}23{]} H. Booth, W. Fisher, R. Galluzzo, and J. Roberts. Accelerating the adoption of software and artificial intelligence agent identity and authorization. Concept paper, NIST National Cybersecurity Center of Excellence, February 2026. \url{https://www.nccoe.nist.gov}

{[}24{]} National Institute of Standards and Technology. AI Agent Standards Initiative. 2026. \url{https://www.nist.gov/artificial-intelligence/ai-agent-standards-initiative}

{[}25{]} OWASP GenAI Security Project. Multi-agentic system threat modeling guide v1.0. April 2025. \url{https://genai.owasp.org/resource/multi-agentic-system-threat-modeling-guide-v1-0/}

{[}26{]} World Wide Web Consortium. Verifiable credentials data model v2.0. W3C Recommendation, May 2025. \url{https://www.w3.org/TR/vc-data-model-2.0/}

{[}27{]} M. Mitchell, S. Wu, A. Zaldivar, et al. Model cards for model reporting. In \emph{Proceedings of the Conference on Fairness, Accountability, and Transparency (FAT*)}, pages 220--229, 2019. \url{https://doi.org/10.1145/3287560.3287596}

{[}28{]} Common Criteria for Information Technology Security Evaluation (ISO/IEC 15408). Part 3: Security assurance components. \url{https://www.commoncriteriaportal.org/cc/}

{[}29{]} L. Bainbridge. Ironies of automation. \emph{Automatica}, 19(6):775--779, 1983. \url{https://doi.org/10.1016/0005-1098(83)90046-8}

{[}30{]} J. E. Bahner, A.-D. Hüper, and D. Manzey. Misuse of automated decision aids: Complacency, automation bias and the impact of training experience. \emph{International Journal of Human-Computer Studies}, 66(9):688--699, 2008. \url{https://doi.org/10.1016/j.ijhcs.2008.06.001}

{[}31{]} Stripe. Minions: Stripe's one-shot, end-to-end coding agents. Stripe Dot Dev Blog, February 2026. \url{https://stripe.dev/blog/minions-stripes-one-shot-end-to-end-coding-agents}

{[}32{]} Stripe. Minions: Stripe's one-shot, end-to-end coding agents---part 2. Stripe Dot Dev Blog, February 2026. \url{https://stripe.dev/blog/minions-stripes-one-shot-end-to-end-coding-agents-part-2}

{[}33{]} H. Son. Goldman Sachs is piloting its first autonomous coder in major AI milestone for Wall Street. CNBC, July 11, 2025. \url{https://www.cnbc.com/2025/07/11/goldman-sachs-autonomous-coder-pilot-marks-major-ai-milestone.html}

{[}34{]} MLCommons Financial Services Working Group. Four-axis option universe: Autonomy \(\times\) ODD \(\times\) action class \(\times\) control envelope. Companion one-pager to the architecture memo, June 2026. Reproduced as Exhibits B1--B3.

{[}35{]} A. Mason, S. Krishnamurthy, and A. Upadhyay. Meet Stripe's knowledge AI platform. Stripe Dot Dev Blog, July 2026. https://stripe.dev/blog/meet-stripes-knowledge-ai-platform

{[}36{]} OpenAI. BNY builds ``AI for everyone, everywhere'' with OpenAI. OpenAI case study, 2026. \url{https://openai.com/index/bny/}

{[}37{]}H. Gold. An OpenAI test model escaped and broke into a real company's server. CNN, July 22, 2026. \url{https://www.cnn.com/2026/07/22/tech/openai-hugging-face-ai-cybersecurity}

{[}38{]} M. Arnold, R. K. E. Bellamy, M. Hind, S. Houde, S. Mehta, A. Mojsilovic, R. Nair, K. Natesan Ramamurthy, D. Reimer, A. Olteanu, D. Piorkowski, J. Tsay and K. R Varshney.. "\href{https://doi.org/10.1147/JRD.2019.2942288}{FactSheets: Increasing Trust in AI Services through Supplier\textquotesingle s Declarations of Conformity}," \emph{IBM Journal of Research and Development}, pp. 6:1-6:13, 2019

{[}39{]} J. A. Piepszak. Letter to shareholders. JPMorganChase 2025 Annual Report, April 6, 2026. \href{https://www.jpmorganchase.com/ir/annual-report/2025/ar-ceo-letter-jennifer-piepszak}{\ul{JPMorganChase annual repor}}t

{[}40{]} DBS Bank Ltd. DBS scales agentic AI to transform way of working for corporate bankers, freeing up time for more strategic client engagements. News release, August 19, 2026. \href{https://www.dbs.com/NewsPrinter.page?locale=en\&newsId=msd3p9eh}{\ul{DBS announcement}}

{[}41{]} Stripe. Minions have changed software engineering at Stripe. LinkedIn post, February 2026. \href{https://www.linkedin.com/posts/stripe_minions-have-changed-software-engineering-activity-7430340204831608832-xC1H}{\ul{Stripe company post}}

\end{document}